\documentclass[reprint,amsmath,amssymb,aps,prb,superscriptaddress,longbibliography]{revtex4-1}

\pdfoutput=1

\usepackage{amsmath,amssymb,mathtools,amsfonts}
\usepackage{graphicx, dcolumn, float, placeins}
\usepackage[version=4]{mhchem}
\usepackage{url}
\usepackage{mciteplus}
\usepackage{comment}
\usepackage{xr}
\usepackage{pdfpages}
\usepackage{pgffor}
\usepackage[utf8]{inputenc}
\DeclareUnicodeCharacter{22EF}{\ldots}
\usepackage{lineno}
\usepackage{microtype}

\usepackage{hyperref}
\hypersetup{breaklinks,colorlinks=true,linkcolor=blue,citecolor=blue,urlcolor=blue,filecolor=blue}

\usepackage{cleveref}
\makeatletter
\AtBeginDocument{\let\LS@rot\@undefined}
\makeatother

\newcommand\mat\mathbf

\DeclareUnicodeCharacter{22EF}{\ldots}

\newcommand{\bytedanceus}{\affiliation{ByteDance Seed, San Jose, CA 95110, United States}}
\newcommand{\bytedance}{\affiliation{ByteDance Seed, Beijing 100098, People’s Republic of China}}

\newcommand{\pekingphysics}{\affiliation{School of Physics, Peking University, Beijing 100871, People’s Republic of China}}
\newcommand{\pekingcenter}{\affiliation{Interdisciplinary Institute of Light-Element Quantum Materials, Frontiers Science Center for Nano-Optoelectronics, State Key Laboratory of Artificial Microstructure and Mesoscopic Physics, Peking University, Beijing 100871, People’s Republic of China}}
\newcommand{\equalcontribution}{\affiliation{These authors contributed equally to this work and are listed in alphabetical order}}

\begin{document}
\author{Changsu Cao}
\email{caochangsu@bytedance.com}
\equalcontribution
\bytedance

\author{Hung Q. Pham}
\email{hung.pham@bytedance.com}
\equalcontribution
\bytedanceus

\author{Zhen Guo}
\bytedance

\author{Yutan Zhang}
\altaffiliation{Current address: Department of Physics and Astronomy, University of California, Davis, California 95616, United States.}
\bytedanceus

\author{Zigeng Huang}
\bytedance

\author{Xuelan Wen}
\bytedance

\author{Ji Chen}
\email{ji.chen@pku.edu.cn}
\pekingphysics
\pekingcenter

\author{Dingshun Lv}
\email{lvdingshun@bytedance.com}
\bytedance

\title{Quantum Many-Body Simulations of Catalytic Metal Surfaces}

\begin{abstract}
Quantum simulations of metal surfaces are critical for catalytic innovation. Yet existing methods face a cost-accuracy dilemma: density functional theory is efficient but system-dependent in accuracy, while wavefunction-based theories are accurate but prohibitively costly. Here we introduce FEMION~(Fragment Embedding for Metals and Insulators with On-site and Nonlocal Correlation), a systematically improvable quantum embedding framework that resolves this challenge by capturing partially filled electronic states in metals. FEMION combines auxiliary-field quantum Monte Carlo for local catalytic sites with a global random-phase approximation treatment of nonlocal screening, yielding a scalable approach across diverse catalytic systems. Using FEMION, we address two long-standing challenges: determining the preferred CO adsorption site and quantifying the \ce{H2} desorption barrier on Cu(111). Furthermore, our calculations demonstrate that the recently discovered 10-electron-count rule can also be extended to single-atom catalysis processes on 3d metal surfaces, resolving controversies arising from density functional theory calculations. We thus open a predictive, first-principles route to modeling complex catalytic systems.
\end{abstract}

\maketitle
\newpage

\section{Main}
Quantitative prediction of chemical reactions on metal surfaces remains a long-standing challenge in computational catalysis~\cite{doi:10.1126/science.adq0102, doi:10.1021/acs.chemrev.7b00776, Vogt2022ActiveSite,D4CS00768A, doi:10.1021/acs.chemrev.0c01060,norskovComputationalDesign2009}. Density functional theory~(DFT)~\cite{hohenberg1964inhomogeneous,kohn1965self} is the dominant method due to its efficiency. However, its inherent flaws often yield quantitatively inaccurate or even qualitatively incorrect predictions for crucial catalytic properties, such as reaction barriers and adsorption sites~\cite{doi:10.1021/acscatal.8b02793}. Wavefunction theory~(WFT), represented by the ``gold standard" CCSD(T) and its variants~\cite{CCSDct, PhysRevLett.131.186402}, offers a systematic path to chemical accuracy. Nevertheless, its prohibitive computational scaling makes the application of WFT to realistic catalytic surfaces impractical.

In practice, the random-phase approximation~(RPA)~\cite{renRandomphaseApproximation2012,RPA} has established itself as a state-of-the-art method in computational materials science, bridging the gap between affordable DFT and prohibitively expensive WFT. RPA plays a dual role: it is often viewed as the highest rung of ``Jacob's ladder" of DFT~\cite{perdewJacobsLadder2001} and, at the same time, as the lowest-level coupled-cluster approximation within WFT~\cite{scuseriaGroundState2008}. This dual character enables RPA to capture long-range dynamical correlation effects, including dispersion and screening, that are absent in standard functionals, thereby correcting qualitative failures of DFT in surface adsorption~\cite{Schimka2010_mbpt_surface_adsorption} and in challenging reactions such as the~\ce{CO2} reduction reaction~\cite{chengElucidatingProton2025,weiImprovingAccuracy2022,weiModelingElectrochemical2022,oudotReactionBarriers2024,szaroBenchmarkingAccuracy2023}. 
At the same time, RPA remains limited, as one of the simplest wavefunction methods, it does not include the essential static correlation needed for accurate descriptions of bond breaking and formation at catalytic active sites. These limitations highlight the urgent need for methods that can capture both long-range screening effects and many-body interactions at catalytic sites. Such an approach would play a pivotal role in shaping the next stage of predictive computational catalysis. 

Balancing accuracy with computational cost remains a central challenge in quantum chemistry development~\cite{shiNatChem2025,shiJACS2023}. Quantum embedding provides a solution by applying high-level solvers to chemically relevant fragments and using lower-cost methods for the surrounding environment~\cite{sunQuantumEmbedding2016}. Catalytic surfaces are a natural application for quantum embedding: the chemically active sites demand accurate many-body treatments, while the extended metallic or insulating environment requires lower-cost approximations. Existing quantum embedding approaches for catalysis have largely followed two distinct paths, each with critical limitations. Density-based methods, such as density functional embedding theory~(DFET), have been applied to metallic systems for decades~\cite{DFET1,DFET2} but they struggle with non-metallic materials, where partitioning across covalent and ionic bonds results in unphysical fragments and uncontrollable errors. 
In contrast, density-matrix-based embedding methods, such as density matrix embedding theory~(DMET)~\cite{DMET_Knizia1,DMET_Knizia2,DMET_Wouters,pDMET_Hung}, systematically improvable embedding~(SIE)~\cite{SIE_booth,SIE} and local natural orbitals~(LNO)~\cite{rolikGeneralorderLocal2011,yePeriodicLocal2024a}, have shown great success for systems with finite band gaps, ranging from molecules to strongly correlated materials such as cuprates, as well as extended insulating surfaces~\cite{cuiSystematicElectronic2022a,Regional_emb}. However, they fail to capture the physics of partially filled electronic states, \textit{i.e.} fractional orbital occupations, that arise in metals with vanishing band gaps. To date, a unified framework that can treat both metallic and insulating systems with high accuracy and on equal footing remains an open challenge.

\begin{figure*}[ht]
 \centering
 \includegraphics[width=\linewidth]{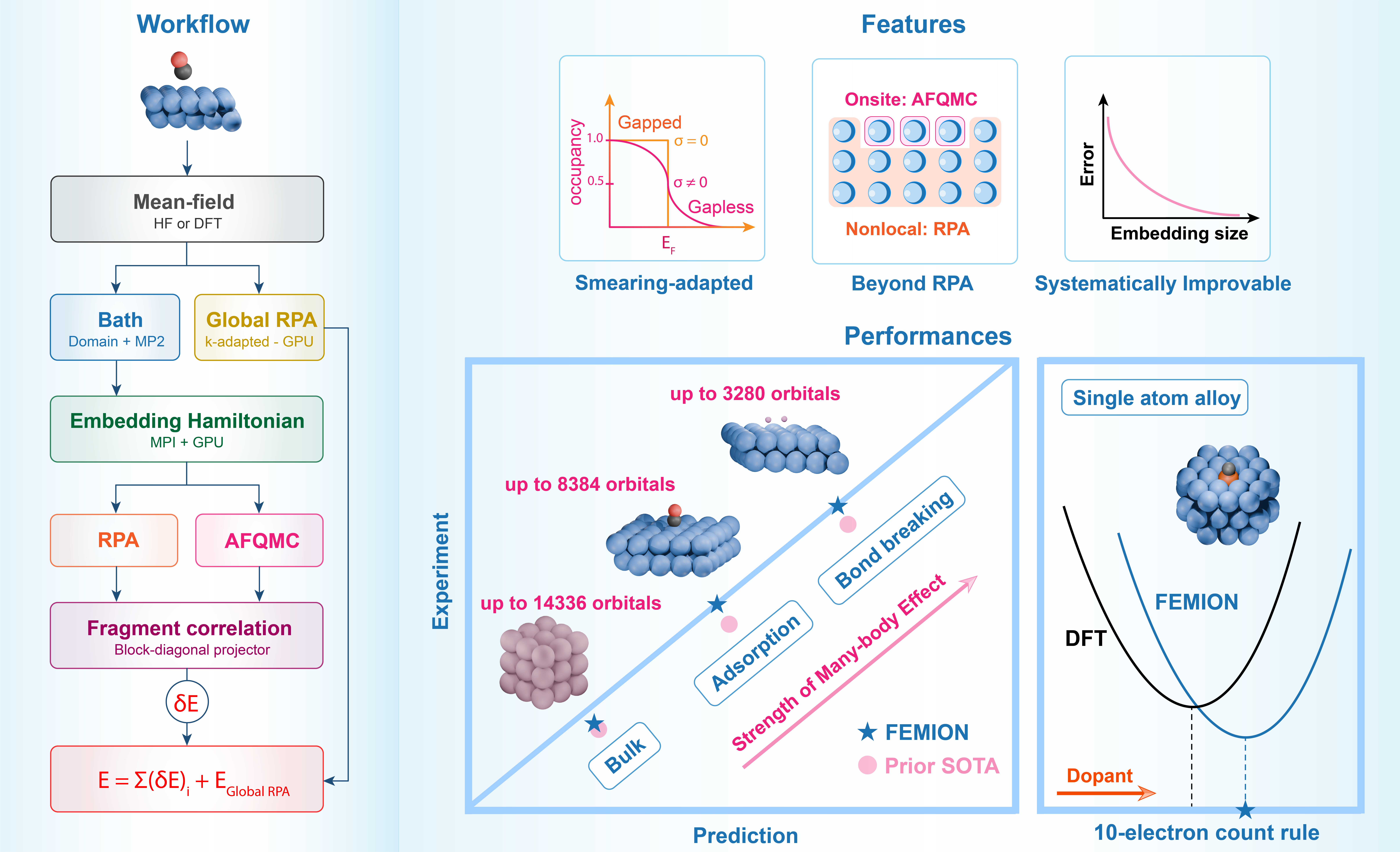}
 \caption{\textbf{FEMION workflow, features, and performance.} The framework combines global RPA with fragment AFQMC (workflow), incorporates smearing adaptation, beyond-RPA accuracy, and systematic improvability (features), and achieves state-of-the-art performance from bulk to adsorption and bond breaking, surpassing prior methods (performance). }
 \label{fig:femion}
\end{figure*}

In this work, we present Fragment Embedding for Metals and Insulators with On-site and Nonlocal Correlation~(FEMION), the first embedding framework designed to capture both the long-range screening effects of extended environments and the many-body correlations at catalytic active sites, treating metallic and insulating systems on equal footing and with high accuracy~(Fig.~\ref{fig:femion}). 
Building on the systematically improvable philosophy underlying SIE and LNO, FEMION establishes a conceptual advance by unifying the treatment of metallic and insulating systems within a single framework. This is achieved by consistently addressing partially filled electronic states throughout bath construction, fragment projection, and solver adaptation, thereby overcoming a long-standing limitation in quantum embedding. Notably, FEMION delivers accuracy beyond RPA by embedding an auxiliary-field quantum Monte Carlo~(AFQMC) solver~\cite{doi:10.1021/acs.jctc.2c00802, maloneIpiePythonBased2023,jiangImprovedModularity2024, huang2024gpuacceleratedauxiliaryfieldquantummonte,mottaInitioComputationsMolecular2018,zhangQuantumMonteCarlo2003a}, a promising and flexible many-body method whose accuracy can systematically improve with better trial wavefunctions, within a global RPA framework that captures the nonlocal screening of the extended environment. To make large-scale simulations feasible, FEMION is fully GPU-accelerated. In our largest calculation, the global RPA reaches supercells with 576 copper atoms, corresponding to roughly 17,000 basis functions, an unprecedented scale for this level of theory.

We validate the efficiency and accuracy of FEMION using several notoriously challenging benchmarks: cohesive energies of bulk Li and Al, CO adsorption on Cu(111), and~\ce{H2} desorption from Cu(111). Our results achieve chemical accuracy while maintaining robustness in metallic systems, where conventional approaches and previous embedding methods often lack intrinsic accuracy or suffer from numerical instability~\cite{COPtpuzzle,cohenInsightsCurrent2008a,sunQuantumEmbedding2016}. Beyond these benchmarks, we apply FEMION to resolve a recently reported conundrum: the apparent violation of the ten-electron count rule in 3d transition-metal-doped single-atom alloys (SAAs)~\cite{schumannTenelectronCountRule2024}. While standard DFT captures the overall trend, its inadequate description of electron correlation shifts the minimum away from the expected ten-electron value. Our many-body treatment corrects this discrepancy, recovering the rule in quantitative detail. This work thus establishes a scalable and systematically improvable path for the \textit{ab initio} modeling of complex catalytic reactions.

\section{Results}\label{sec:results}

\subsection*{Methodological Advancements}

FEMION is a systematically improvable quantum embedding framework designed to resolve the challenge of accurately capturing local many-body correlations while maintaining scalability for extended systems, such as catalytic metal surfaces~(Fig.~\ref{fig:femion}). 
FEMION achieves this through two key methodological advances 

First, FEMION employs a domain-localized bath construction that restricts the environment subspace to the physically relevant region around each fragment. This design circumvents the need to compute natural orbitals for the full supercell using expensive wavefunction methods (e.g., MP2), effectively decoupling the correlation problem from the scale of Brillouin-zone sampling. As a result, the method ensures scalability to systems with thousands of orbitals, enabling large metallic supercells that are essential for capturing long-range screening and surface reconstructions.

Second, and more critically, FEMION explicitly treats the gapless electronic states that are inherent to metallic systems. By introducing thermal smearing as a fictitious temperature, FEMION eliminates the numerical divergences that are often encountered in gapless metallic systems. We generalize both the RPA and phaseless AFQMC solvers to work with the resulting fractional occupations, ensuring robust performance for metals and insulators alike. A key feature of this implementation is the use of a block-diagonal projector, which rigorously defines fragment energies within the smeared environment and links the local embedding problem to the global system. This allows local AFQMC corrections to be added to the global RPA baseline, combining accurate long-range screening with strong local correlation, thus yielding a systematically improvable description of both local and nonlocal correlation effects. The performance panel of Fig.~\ref{fig:femion} illustrates how this combination of domain-localized baths, smearing-adapted solvers, and fragment corrections achieves state-of-the-art accuracy across diverse materials problems, from bulk properties to adsorption and bond breaking at metal surfaces. The workflow panel of Figure~\ref{fig:femion} provides a schematic overview of how these two advances are integrated within the FEMION framework. Full algorithmic details are provided in Section~\ref{sec:methods} and the Supplementary Information.

\subsection*{Systematic Improvability Across Gapped and Gapless Mean Field}

\begin{figure*}[ht]
 \centering
 \includegraphics[width=\linewidth]{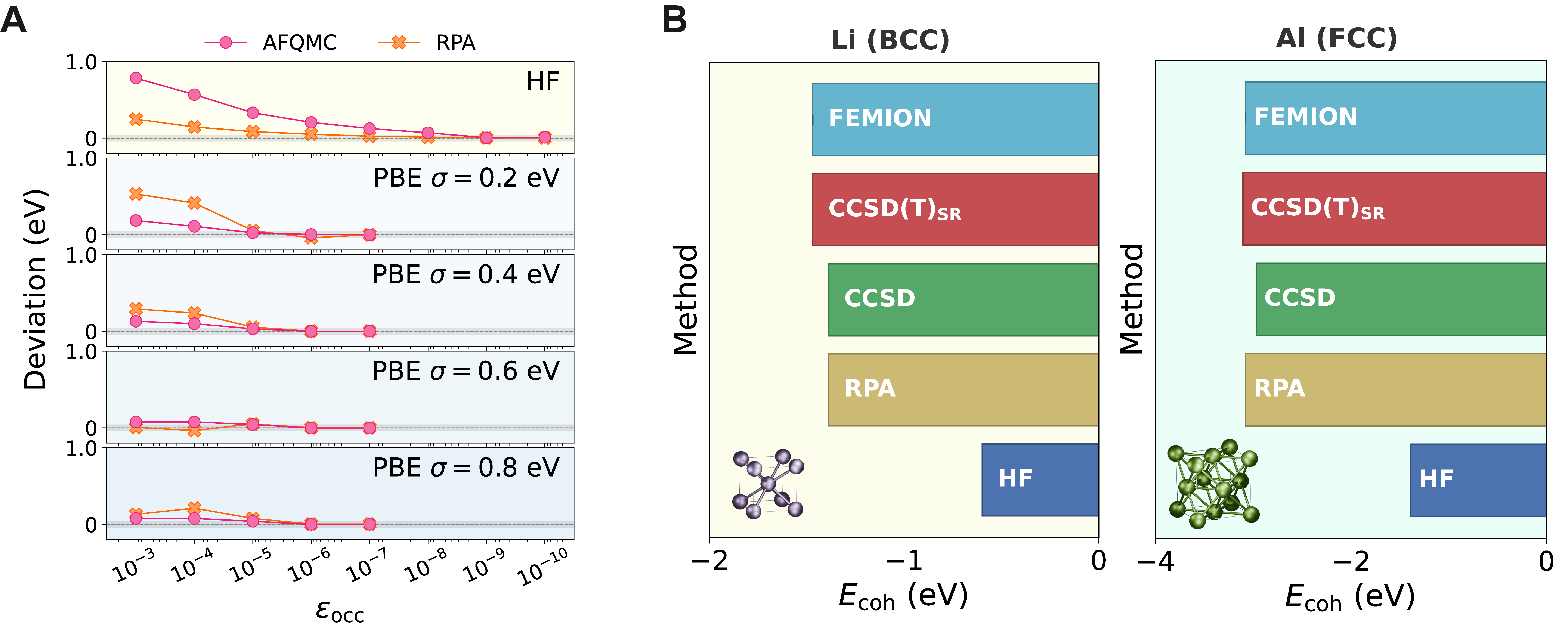}
 \caption{\textbf{Benchmarking the Accuracy of the FEMION Embedding Scheme on Pure Metals.} \textbf{(A)} Deviations of embedding RPA and AFQMC total energy from their corresponding conventional counterparts, using HF and PBE~($\sigma = 0.2-0.8$ eV) as mean-field wavefunctions, decrease with tighter $\varepsilon_\mathrm{occ}$ and consistently approach the conventional limit, with only weak dependence on the smearing parameter $\sigma$. \textbf{(B)} Cohesive energies for simple metallic systems: lithium in the body-centered cubic~(BCC) structure and aluminum in the face-centered cubic~(FCC) structure, demonstrating agreement with high-level theory. The detailed reference values are listed in Table~\ref{SI-tab:li_al_properties} of the Supplementary Information.}
 \label{fig:bulk}
\end{figure*}

A desirable feature of any robust quantum embedding or local quantum chemical method is the ability to systematically approach the exact full-system limit by tightening controlled thresholds. This behavior has been demonstrated in molecular systems with gapped mean-field references, as in SIE and LNO. However, it has not yet been demonstrated in systems with metallic character~\cite{SIE, SIE_booth,yePeriodicLocal2024a,kurianLinearScaling2024}. To address this gap, in Fig.~\ref{fig:bulk}A, we benchmark the absolute total embedding energy of lithium against conventional RPA and AFQMC to illustrate the systematic improvability of FEMION. Both HF and PBE references are considered: HF yields a finite gap, while PBE produces a gapless spectrum, which we probe with different smearing parameters ($\sigma = 0.2$–$0.8$ eV). In all cases, tightening the bath truncation threshold $\epsilon_{\text{occ}}$ results in smooth convergence of the embedded correlation energy toward the reference. Notably, this controlled convergence, long established in insulating systems, is preserved in the challenging metallic regime, providing direct evidence that FEMION achieves systematically improvable embedding for metals.

In addition, we assess the effect of different virtual-to-occupied cutoff ratios ($\epsilon_{\text{vir}}/\epsilon_{\text{occ}}$) using the RPA solver, as shown in Fig.~\ref{SI-fig:threshold_testing} of the Supplementary Information. This parameter controls how many virtual orbitals are retained relative to the occupied-space threshold used in the embedding. We find that a ratio of 10 achieves the same convergence rate as a more aggressive ratio of 100, while avoiding the unnecessary inclusion of an excessive number of virtual orbitals in the active space. This choice therefore provides a balanced compromise between accuracy and computational efficiency, and we adopt $\epsilon_{\text{vir}}/\epsilon_{\text{occ}}=10$ for all subsequent calculations. \\

\textbf{Cohesive Energy of Simple Metals} \\

We begin by evaluating the performance of our framework on two elemental metals, lithium~(BCC) and aluminum~(FCC), both widely used benchmarks with well-characterized cohesive energies. Although pure metals are systems where semilocal DFT functionals (e.g., PBE~\cite{PBE}) often perform well, they remain notoriously challenging for high-level wavefunction methods due to their metallic character. This makes them valuable test cases for assessing novel wavefunction-based embedding approaches~\cite{Verena_LiAl,mihmShortcutThermodynamic2021,CCSDct,yePeriodicLocal2024a,zhangPerformanceVarious2018}. 

As shown in Fig.~\ref{fig:bulk}B, HF provides the mean-field baseline, and its severe underestimation of cohesive energies in both Li and Al underscores the critical role of many-body correlation treatments. RPA systematically improves upon this by incorporating nonlocal screening, but still underbinds relative to higher-level methods. CCSD recovers much of the missing correlation but still deviates from CCSD(T)\textsubscript{SR}~\cite{Verena_LiAl}. The latter mitigates infrared divergences in metallic systems and, while approximate, represents the most reliable computational benchmark currently available for both Li and Al. We note that a recent work of CCSD(cT) also shows high accuracy on uniform electron gas and Li~\cite{CCSDct}. FEMION yields cohesive energies in close agreement with CCSD(T)\textsubscript{SR}, demonstrating that it can reliably capture correlation effects in extended metallic systems. Crucially, unlike conventional wavefunction approaches tied to an HF reference, FEMION can start from either HF or DFT reference. For metals, a PBE reference provides a more accurate mean-field reference and, together with FEMION’s fractional-occupation handling, leads to smoother orbital occupations and stable convergence. This flexibility is essential for extending FEMION to larger and more complex materials, as explored in the following sections.\\

\textbf{CO Adsorption on the Cu(111) Surface} \\

\begin{figure*}[ht]
 \centering
 \includegraphics[width=\linewidth]{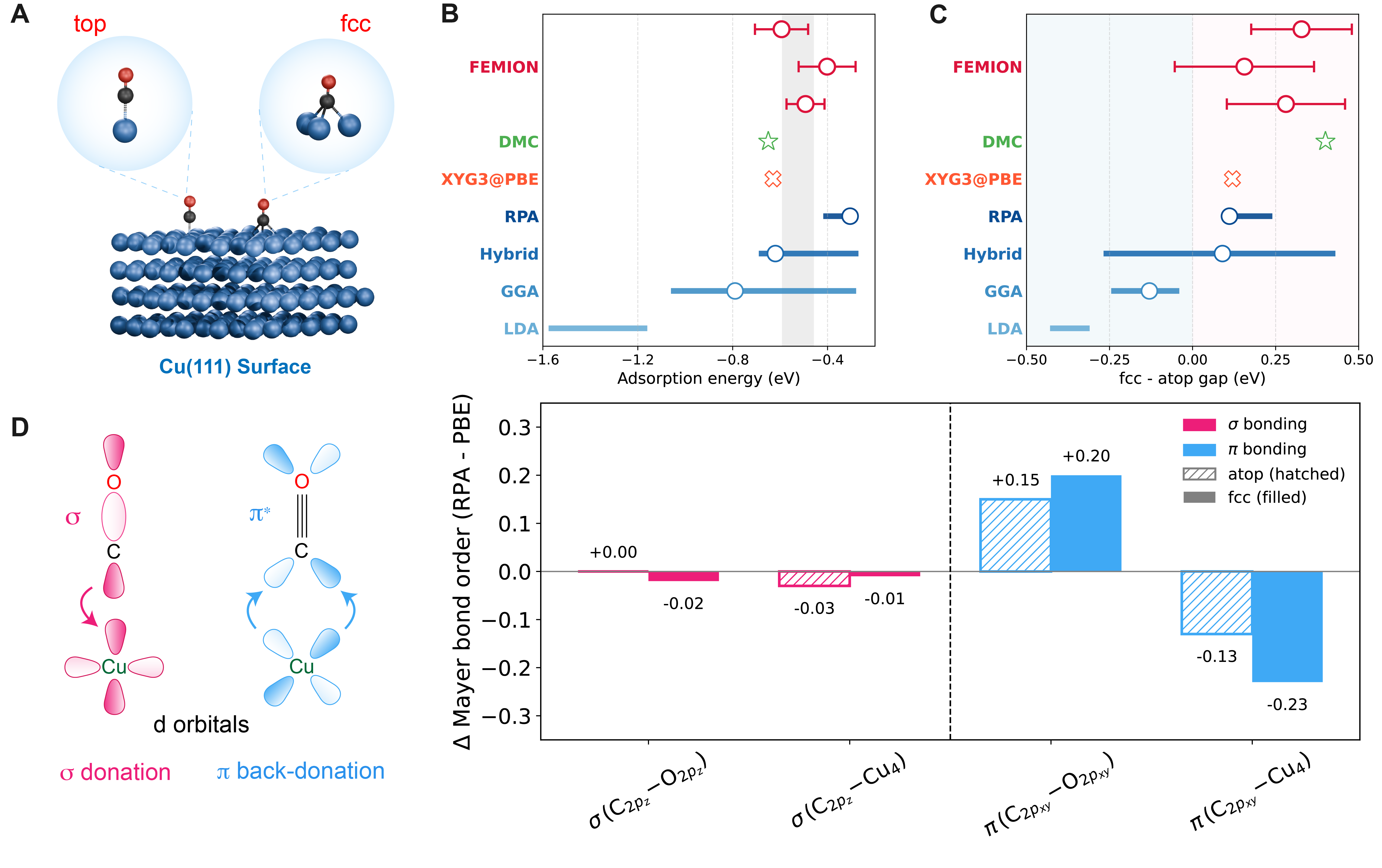}
 \caption{\textbf{Application of the FEMION Framework to CO Adsorption on Cu(111)}. \textbf{(A)} Atop and fcc configurations for the CO@Cu(111) system. 
\textbf{(B)} Absolute adsorption energies at the atop site, with our results~(empty circles) compared against a range of methods from DFT to QMC. Detailed definitions for the DFT methods are provided in Table~\ref{SI-tab:COCu_data}. FEMION results are shown at three computational settings~(from bottom to top):~(1) 3×3 $k$-mesh with $\varepsilon_{\text{occ}} = 0.005$ and $\varepsilon_{\text{vir}} = 0.0005$;~(2) 3×3 $k$-mesh with $\varepsilon_{\text{occ}} = 0.005$ and $\varepsilon_{\text{vir}} = 0.00005$; and~(3) 4×4 $k$-mesh with $\varepsilon_{\text{occ}} = 0.005$ and $\varepsilon_{\text{vir}} = 0.0005$. The gray band indicates the experimental range.
\textbf{(C)} Energy difference between fcc and atop configurations using the same methods as in panel~(B); a positive value indicates correct preference for the atop site, consistent with experiment. 
\textbf{(D)} Left: schematic of the two dominant interactions between CO and Cu: $\sigma$ donation and $\pi$ back-donation.~\cite{housecroftInorganicChemistry2005,hieringerBlyholderModel2020} Right: bar plots of Mayer bond orders for these interactions~(C–O and C–Cu) in both atop and fcc configurations. The experimental value is corrected for ZPE and detailed reference values are listed in Table~\ref{SI-tab:COCu_data} of the Supplementary Information.}
 \label{fig:COCu}
\end{figure*}

Accurate theoretical modeling of chemisorption is critical for understanding catalytic processes, but has long faced persistent challenges. A well-known example is the ``CO puzzle", in which conventional DFT methods fail qualitatively by predicting the wrong adsorption site, and quantitatively by overestimating the CO binding energy on transition metal surfaces~\cite{COPtpuzzle}. This issue is especially critical for copper, an important catalyst for $\rm{CO_{2}}$ reduction to sustainable fuels and chemical feedstocks~\cite{araujoAdsorptionEnergies2022,chenAccurateDescriptions2023,Schimka2010_mbpt_surface_adsorption}. 

To further evaluate the performance of our framework in chemically realistic scenarios, we apply FEMION to CO@Cu(111). We examine two prototypical CO adsorption sites on Cu(111) (Fig.~\ref{fig:COCu}A). The atop site, where CO binds above a surface Cu atom, has been established experimentally as the most favorable geometry for this system. The fcc hollow site, a threefold hollow aligned with a second-layer Cu atom, is often incorrectly predicted by semilocal DFT to be preferred. We focus on these two sites, as they are the most extensively characterized in both experiments and computations, serving as the standard benchmarks for Cu(111). This failure of DFT stems from its tendency to place the CO lowest unoccupied molecular orbital~($\pi^*$ orbital) at artificially low energy, leading to an overestimation of metal-to-ligand $\pi$ back-donation. This, in turn, incorrectly stabilizes multicoordinated sites such as the fcc. Therefore, CO@Cu(111) provides an ideal system for assessing whether an embedding method can properly balance localized molecular states and delocalized metallic screening.

Fig.~\ref{fig:COCu}B compares our computed CO adsorption energy on the atop site of Cu(111) with experimental data and state-of-the-art theoretical methods. DFT functionals across LDA, GGA, and hybrid classes show a general trend of improving accuracy from LDA to GGA to hybrids, yet reported values in the literature remain highly scattered, spanning a wide range. As expected, LDA consistently overbinds, while hybrids tend to perform better, though their accuracy remains inconsistent across systems. RPA offers a more reliable baseline by reducing the variability seen in DFT predictions, yet systematically underestimates CO binding on Cu(111).

Notably, our embedding method yields adsorption energies that not only fall within the experimental range, but also achieve excellent agreement with the state-of-the-art diffusion Monte Carlo~(DMC). While recent results from embedded double-hybrid functionals like XYG3@PBE~\cite{chenAccurateDescriptions2023} also show good agreement with benchmarks, our approach is fundamentally different: it evaluates correlated energies directly from many-body wavefunctions rather than from density functionals. In contrast to DFT-based embedding schemes, where the final energy depends on the functional, our method uses DFT only to generate orbitals; the correlated energy is obtained solely from AFQMC and RPA, free from any DFT exchange–correlation contributions. This guarantees that our framework remains rooted in a first-principles many-body treatment, even when initialized with DFT orbitals.

Our prediction of the preferred adsorption site, based on the computed energy difference between the fcc and atop configurations (Fig.~\ref{fig:COCu}C), is consistent with both DMC and XYG3@PBE, all favoring the atop site. Interestingly, hybrid functionals show inconsistent behavior: depending on the specific functional and computational setup, they can favor either the atop or fcc site. For the absolute value of the site preference gap, RPA predicts the correct site with values in the range of 0.10–0.24 eV, smaller than the DMC benchmark of about 0.40 eV. XYG3@PBE yields a gap of 0.12 eV. FEMION gives a gap between 0.16–0.33 eV, slightly larger than RPA and with the upper end approaching the DMC value. The relatively large error bars in FEMION reflect error propagation from subtracting two adsorption energies that each carry their own uncertainty; despite this, the method produces smooth, stable results that remain consistent with higher-level benchmarks. \\


To understand the bonding mechanism of CO adsorption on the Cu(111) surface, we examine changes in Mayer bond orders between mean-field PBE and many-body RPA~($\Delta$ = RPA $-$ PBE), as shown in Fig.~\ref{fig:COCu}D. A first key observation is that the $\sigma$-bond order involving C(2$p_z$)–O(2$p_z$) and C(2$p_z$)–Cu remains largely unaffected by the nonlocal many-body effects captured in RPA, indicating that $\sigma$ donation is insensitive to electron correlation beyond DFT. In contrast, the $\pi$-bonding components exhibit pronounced correlation effects. Under RPA, the bond order of the $\pi$ bond between C(2$p_{xy}$) and O(2$p_{xy}$) increases significantly, indicating a strengthening of the C–O bond. In contrast, the $\pi$-backbonding between Cu(3d) and C(2$p_{xy}$) is substantially reduced, reflecting a weaker interaction between the adsorbate and the metal surface. Importantly, these opposing trends are more pronounced at the bridge~(fcc) site than at the atop site: the C–O bond strengthens more and the C–Cu bond weakens more in the fcc configuration. This qualitative signature of stronger internal bonding and weaker metal–adsorbate interaction aligns with quantitative many-body predictions that the fcc site is less favorable for CO adsorption, with a higher binding energy relative to the atop site. This selective weakening of $\pi$ back-donation corrects the well-known overbinding tendency of DFT, particularly at high-coordination sites. By including nonlocal correlation, RPA mitigates the delocalization error in approximate DFT functionals and shifts the $\pi^*$ orbital to higher energy~\cite{Schimka2010_mbpt_surface_adsorption}, reducing spurious metal-to-ligand back-donation. Collectively, these effects provide a consistent physical explanation for why RPA, and similarly FEMION,  resolves the CO adsorption site preference problem.

Together, these results demonstrate that many-body effects not only enhance energetic accuracy but also reshape the underlying chemical bonding picture, leading to a more physically realistic and experimentally consistent description of adsorption site preference, exemplified here by CO on metal (111) surfaces. \\

\textbf{Reaction Barrier for \ce{H2} Desorption from Cu(111)} \\

\begin{figure}[ht]
 \centering
 \includegraphics[width=\linewidth]{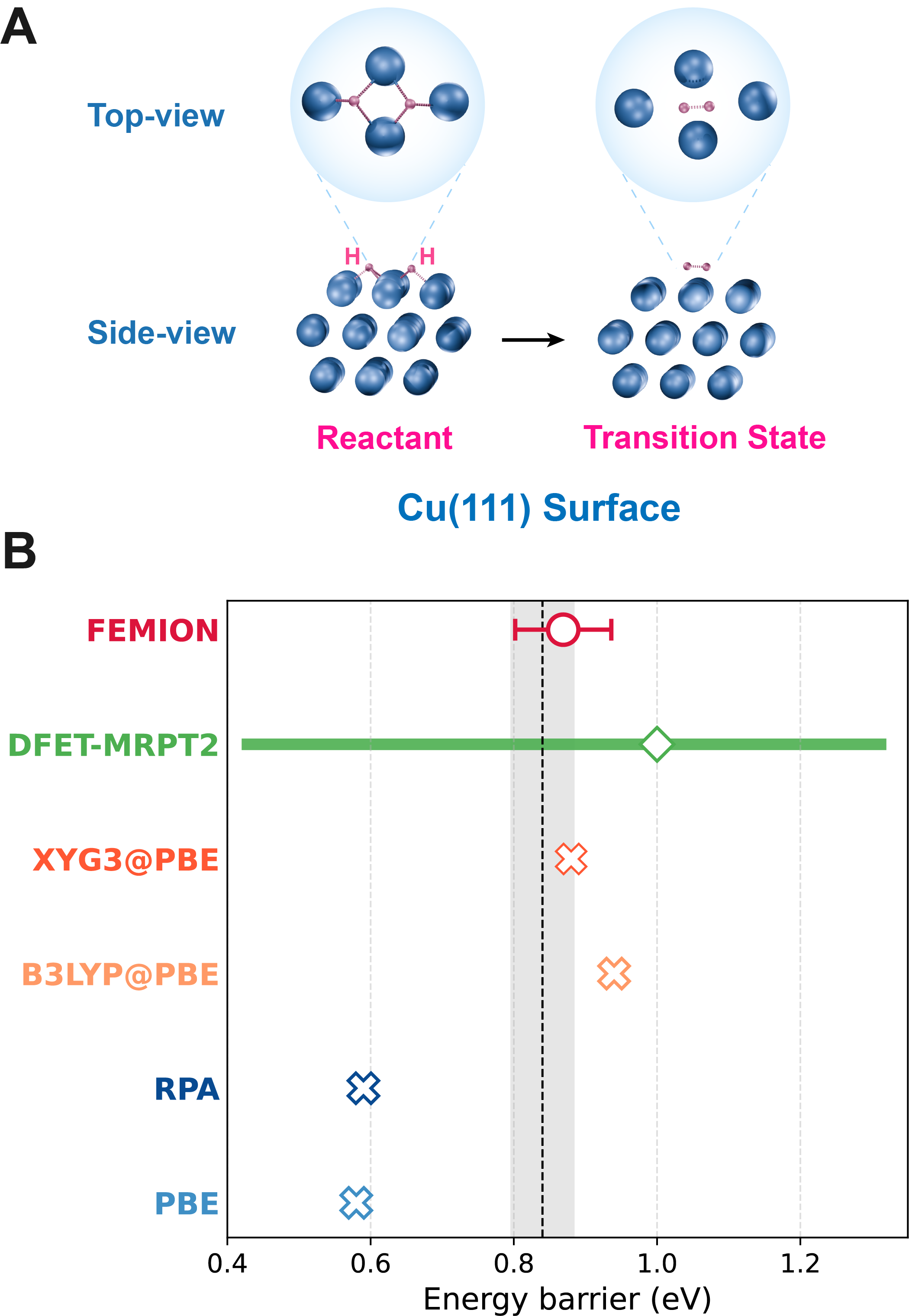}
 \caption{\textbf{Application of the FEMION Framework to \ce{H2} Dissociation on Cu(111)}. \textbf{(A)} Schematic of the reaction pathway from the adsorbed reactant state to the transition state for \ce{H2} dissociation on the Cu(111) surface. \textbf{(B)} Calculated energy barriers~(bottom) using various methods, including the FEMION embedding, DFET-MRPT2, DFT-based cluster embedding~(XYG3@PBE, B3LYP@PBE), RPA, and PBE. The red circle represents our FEMION result. The green bar illustrates the range of energy barriers predicted by different multireference solvers within the DFET-MRPT2 framework, with the diamond marking the specific result that agrees best with experiment. Crosses represent reference values from the literature. The dashed line indicates the experimental barrier, with the gray band representing the chemical accuracy of 1 kcal/mol. The experimental value is corrected for zero-point energy~(ZPE) and detailed reference values are listed in Table~\ref{SI-tab:H2Cu_data}.} 
 \label{fig:H2Cu}
\end{figure}

The desorption of \ce{H2} from Cu(111) is another prototypical yet challenging benchmark for electronic structure methods applied to metallic systems~(see Fig.~\ref{fig:H2Cu}A). The accurate calculation of a reaction energy barrier requires a balanced treatment of both the reactant and the corresponding transition state. Transition-state structures, with stretched or partially broken bonds, typically exhibit stronger static electron correlation than their reactant counterparts at equilibrium geometries~\cite{Delocalization2008,cohenInsightsCurrent2008a}. As a result, DFT often underestimates reaction barriers severely. To avoid systematic bias in the predicted barrier heights, an electronic structure method must capture this correlation energy accurately~\cite{zhaoDesignDensity2006}. Here, we compare our embedding framework with several state-of-the-art methods. 

Conventional single-reference methods struggle in this regime. As shown in Fig.~\ref{fig:H2Cu}B, this is evident in the poor performance of standard DFT with the PBE functional and post-mean-field approaches such as RPA, both of which significantly underestimate the reaction barrier. This failure is characteristic of single-reference methods and is often attributed to a combination of delocalization error and the inability to capture static correlation. 
Although density embedding methods have shown improved performance, their application can also present challenges. The embedded DFT approach yields a barrier that slightly overestimates the reference value~(Fig.~\ref{fig:H2Cu}B). Furthermore, their performance can be sensitive to the choice of functional for both the high-level and low-level regions, without a clear path to systematic improvement.

Although one might expect that a high-level multireference solver within an embedding framework would resolve these issues, this is not necessarily the case. For example, DFET combined with embedded multireference second-order perturbation theory~(emb-MRPT2) shows a pronounced dependence on the choice of multireferential solver~\cite{zhaoBenchmarkingEmbedded2020}. 
The calculation using emb-CASPT2~\cite{CASPT2_1, CASPT2_2} yields a barrier of 1.00 eV, in close agreement with experiment, whereas using emb-NEVPT2~\cite{NEVPT2} produces a significantly lower value of 0.64 eV. These discrepancies between formally similar approaches, as illustrated by the green bar in Fig.~\ref{fig:H2Cu}B, underscore the need for high-level solvers that combine high accuracy with consistent robustness to enable predictive catalytic modeling.

In contrast to these challenges, our embedding framework with AFQMC as the high-level solver provides a robust and accurate solution. AFQMC is well-suited for systems with strong correlation, as it effectively balances the description of both static and dynamic correlation effects~\cite{doi:10.1021/acs.jctc.2c00802}. As shown in Fig.~\ref{fig:H2Cu}B, our FEMION framework with the thermal-smearing AFQMC solver yields an energy barrier in good agreement with experiment. This demonstrates that a systematically improvable embedding framework, when paired with AFQMC, a solver that can be further improved with more accurate trial wavefunctions, can overcome the limitations of conventional DFT and embedding schemes, enabling quantitatively reliable predictions for catalytic reactions.

\subsection*{Transition-Metal-Doped Single-Atom Alloys}

\begin{figure*}[ht]
 \centering
 \includegraphics[width=0.85\textwidth]{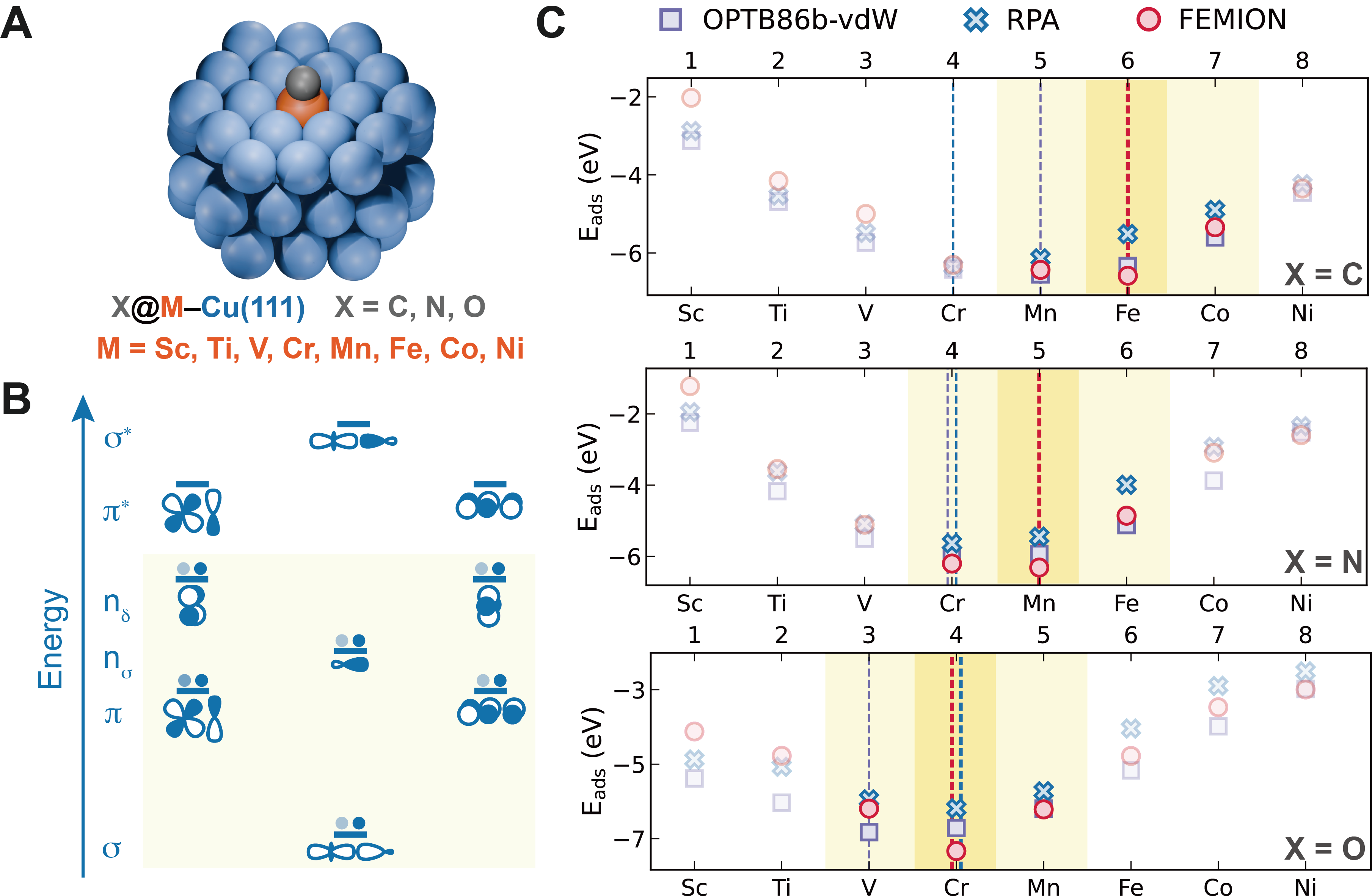}
\caption{\textbf{Application of the FEMION framework to 3d transition-metal–doped single-atom alloys $\text{X}@\text{M–Cu}(111)$.} \textbf{(A)} Structure of $\text{X}@\text{M–Cu}(111)$; \textbf{(B)} Schematic orbital diagram illustrating the ten-electron rule for \(X=\) C, N, O, in which the bonding and non-bonding orbitals are highlighted. \textbf{(C)} Adsorption energies~($E_{\text{ads}}$) as a function of the 3d dopant across different methods: OPTB86b-vdW (purple squares), RPA (blue crosses), and FEMION (red circles). Vertical dashed lines indicate the energy minimum for each computational method. The yellow shaded regions highlight the optimal dopant predicted by the 10-electron count rule and its immediate neighbors to guide the eye, with the optimum itself located in the darker panel with bolder dash line. The small error bars for the FEMION data are omitted for clarity and are presented in Figure~\ref{fig:DFT_range}.}

\label{fig:SSA}
\end{figure*}

Single-atom alloys (SAAs), formed by doping a host metal with isolated atoms (Fig.~\ref{fig:SSA}A), have attracted increasing interest for their unique catalytic properties. 
A recent study proposed a ten-electron count rule to rationalize adsorbate binding trends on dopant atoms in these systems~\cite{schumannTenelectronCountRule2024}. The rule states that binding is strongest within a transition-metal row when the total valence electron count of the dopant and adsorbate fills the bonding and non-bonding orbitals while leaving antibonding orbitals empty, which corresponds to a count of 10 for hydrogen and 12 for p-block adsorbates such as carbon, nitrogen, and oxygen (Fig.~\ref{fig:SSA}B). While this electron-counting picture shows excellent agreement for 4d and 5d transition-metal dopants, the behavior of the 3d series is markedly more complex. This complexity is attributed to strong spin effects in 3d elements. Previous work demonstrated that when spin polarization is artificially suppressed, 3d dopants restore a “V-shaped” adsorption trend similar to that observed for 4d and 5d systems~\cite{schumannTenelectronCountRule2024}.
But a critical discrepancy remains: standard DFT systematically shifts the strongest binding site to the wrong element. In particular, for C, N, and O adsorbates, the binding-strength maximum is shifted to Mn (instead of Fe), Cr (instead of Mn), and V (instead of Cr), respectively, casting doubt on the universality of this chemical rule.

Here, we resolve this conundrum by showing that the discrepancy arises not from the ten-electron rule itself, but from the inadequate treatment of electron correlation in standard DFT. Using the $X@\text{M–Cu}(111)$ SAA as a benchmark (Fig.~\ref{fig:SSA}C), we systematically apply methods of progressively higher levels of theory, from DFT to RPA to FEMION. Overall, our DFT results confirm what has been previously observed: DFT consistently shifts the binding-strength maximum to the left. In particular, for C, baseline DFT places the maximum at Mn ($3d^54s^2$, total valence 11), while RPA shifts it to Cr ($3d^54s^1$, total valence 10). In contrast, FEMION correctly identifies Fe ($3d^64s^2$, total valence 12) as the strongest binding site. For N, both DFT and RPA incorrectly predict Cr as the strongest binder, whereas FEMION recovers Mn at the correct 12-electron count. In the case of O, both RPA and FEMION agree on Cr as the strongest binding site, also corresponding to 12 electrons. Overall, our calculations demonstrate that FEMION systematically recovers the ten-electron rule for 3d single-atom alloys, consistently placing the strongest binding site at the expected electron count, in contrast to both DFT and RPA, which fail to do so. This suggests that, despite capturing nonlocal correlation effects relevant to metallic screening, both DFT and RPA may remain insufficient to fully describe the strong local multireference character associated with partially occupied 3d states. 
To get the qualitative insight into why FEMION resolves the incorrect trend of DFT, we turn to an orbital-level analysis.

The physical origin of this discrepancy is rooted in the electronic nature of the non-bonding frontier orbitals~\cite{schumannTenelectronCountRule2024}. Figure~\ref{fig:SSA}B shows the most relevant frontier states for adsorption, $n_{\delta}$ and $n_{\sigma}$, which originate from dopant 3d states and adsorbate 2p lone pairs that remain non-bonding. These orbitals are highly localized at the top of the valence manifold, and when partially occupied, they exhibit pronounced multireference character. This challenge becomes acute in the 3d series, where the absence of a core d-shell, an effect termed primogenic repulsion, renders the 3d orbitals significantly more contracted than their 4d and 5d counterparts~\cite{caoPhysicalOriginChemical2019}. This contraction amplifies electron--electron repulsion and enhances the multireference character, hence causing the failure of DFT. By explicitly capturing these many-body effects, FEMION reconciles the adsorption behavior of 3d SAAs, establishing a consistent framework for predicting adsorption energetics across the transition metal series.

In addition to the adsorption energy minimum, we observe a plateau-like region emerging near Cr, whose width depends on the adsorbate. On both sides of this region, the adsorption energy decreases and increases more rapidly. This behavior reflects the interplay of two effects. From Sc to Cr before the plateau, stabilization is driven by the progressive lowering and contraction of the 3d states, which improves their energetic and spatial overlapping with the adsorbate valence states and strengthens the metal–adsorbate bond~\cite{schumannTenelectronCountRule2024}. 
This stabilization, however, saturates near Cr. Beyond this point, the adsorption trend is dominated by the sequential filling of the non-bonding and anti-bonding states. 
The occupation of the non-bonding states ($n_{\delta}$, $n_{\sigma}$) only contribute subtly to the bonding interaction, leading to a relatively flat adsorption-energy region that necessitates an accurate description of the electronic structure. FEMION correctly captures the minimum within this plateau, whereas standard DFT methods struggle to do so. Subsequently, the filling of anti-bonding states significantly weakens the metal–adsorbate bond, causing a sharp rise in adsorption energy that terminates the plateau and completes the characteristic V-shaped trend of the adsorption curve.

\section{Discussion}

In this work, we present FEMION, a unified and systematically improvable quantum embedding framework that consistently treats both insulating and metallic systems on equal footing. By combining a global RPA backbone with local embedding RPA for long-range screening and high-level solvers such as AFQMC for chemically active regions, FEMION bridges local and nonlocal correlation within a single formalism. Smearing-adapted projectors and block-diagonal domain definitions ensure numerical stability in gapless systems while retaining accuracy in insulating regimes, thereby extending density-matrix-based embedding beyond its traditional limitations.

Our formulation achieves asymptotic exactness in both weakly and strongly correlated limits: the RPA backbone captures collective metallic fluctuations, while AFQMC corrections recover correlation within chemically active fragments. Benchmark applications demonstrate broad applicability: cohesive energies of elemental metals (Li, Al) are accurately reproduced; CO adsorption on Cu(111) resolves long-standing discrepancies among DFT, RPA, and DMC; the reaction barrier of \ce{H2} desorption from Cu(111) is predicted in excellent agreement with experiment; and the ten-electron rule governing adsorbate binding in 3d single-atom alloys is quantitatively recovered, correcting systematic failures of standard DFT. These results highlight the ability of FEMION to treat localized chemisorption and extended metallic screening in a consistent and scalable manner. 

Looking ahead, continued advances in the development of AFQMC trial wavefunctions, such as multi-determinant expansions, machine-learned~\textit{ansatz}, or other quantum-inspired trial states, will directly benefit our framework. Because FEMION decouples the embedding formulation from the solver backend, these enhancements can be seamlessly integrated to further improve accuracy and efficiency. In particular, better quantum trial wavefunctions~\cite{pham2024scalable, huang2024gpuacceleratedauxiliaryfieldquantummonte} can enhance the treatment of local correlation in complex systems, thereby strengthening the predictive power of fragment-based simulations in catalysis and materials science. This adaptability, coupled with the development of new basis sets tailored for periodic correlated calculations in metals, ensures a robust and evolving future for first-principles simulation. 

\section{Methods}\label{sec:methods}
In this section, we highlight the key steps of the FEMION framework developed in this work. FEMION is designed to bridge the gap between high-level wavefunction theory and extended periodic systems by decoupling local strong correlations from global screening effects. The framework proceeds in three main stages: (1) the construction of a domain-localized correlated bath to reduce the dimensionality of the embedding problem; (2) the generation of a compact, real-valued embedding Hamiltonian from complex $k$-point data; and (3) the solution of this Hamiltonian using a combination of global RPA and local Auxiliary-Field Quantum Monte Carlo (AFQMC). Finally, we describe how these components are synthesized into a total energy formulation that is systematically improvable and robust for metallic systems.

\subsection*{Domain-localized Correlated Bath Construction}

Starting from a standard mean-field calculation (e.g., DFT or Hartree--Fock, with or without thermal smearing), the canonical orbitals are localized to the basis of intrinsic atomic orbital~(IAO) and the projected atomic orbital~(PAO)~\cite{IAO}. The system is then partitioned into fragments based on the localized orbitals, defined either as single atoms or groups of atoms.

The bath orbitals of the fragment are constructed in a two-step manner. First, a DMET~\cite{DMET_Knizia1, DMET_Knizia2, DMET_Wouters} cluster is built by diagonalizing the environment block of the mean-field one-particle density matrix~(1-RDM). For the mean-field with thermal smearing, the 1-RDM is no longer idempotent, and consequently the number of bath orbitals may exceed the number of fragment orbitals. We retain all environmental orbitals with fractional occupations that deviate from fully filled or empty states (occupations 0 or 2) by more than a numerical threshold ($10^{-9}$ in this work).

Subsequently, the bath space is augmented with correlated orbitals derived from approximate MP2 amplitudes, following the Systematically Improvable Embedding (SIE)~\cite{SIE_booth, SIE} approach. While standard SIE is restricted to $\Gamma$-point sampling to ensure real integrals, FEMION extends this formalism to multi-$k$-point sampling. This extension allows for the efficient use of dense $k$-meshes, which is critical for approaching the thermodynamic limit in metallic systems without requiring prohibitively large unit cell.

However, direct application of this bath extension requires treating the entire unentangled environment, the size of which grows rapidly with $k$-point sampling, creating a scalability bottleneck. FEMION resolves this by projecting the unentangled environment onto a Boughton–Pulay (BP) domain~\cite{BPdomain} uniquely defined for each fragment. Because the BP domain is spatially localized, the dimension of the correlated bath is controlled by the local chemical environment rather than the full Brillouin-zone sampling. As a result, the correlated bath subspace remains manageable and systematically improvable, ensuring accuracy while maintaining scalability to the thermodynamic limit. A detailed description of the bath construction procedure is provided in Section~\ref{SI-subsec:domain} of the Supplementary Information.

\subsection*{Embedding Hamiltonian Construction}
Direct computation of the full four-index two-electron integral tensor $(ij|kl)$ is computationally prohibitive for large embeddings due to its $\mathcal{O}(N^4)$ scaling. Furthermore, in periodic systems, momentum-space sampling inherently generates complex-valued electron repulsion integrals (ERIs), which are incompatible with conventional quantum chemical solvers designed for real integrals. To circumvent these limitations, FEMION manipulates the ERIs in their Cholesky-decomposed form (CDERIs)~\cite{RSDF}, transforming them directly from the $k$-point atomic orbital ($k$-AO) basis to the real-space embedding orbital (R-EO) basis.

The CDERIs are initially generated during the mean-field calculation, where the two-electron integrals are approximated as a sum of outer products of Cholesky vectors $L_{\mu\nu}^J$:
\begin{equation}
(\mu\nu|\rho\sigma) \approx \sum_J^{N_{\text{aux}}} L_{\mu\nu}^J [L_{\rho\sigma}^J]^*
\end{equation}
where $\mu,\nu,\rho,\sigma$ denote AOs and $J$ indexes the auxiliary basis of dimension $N_{\text{aux}}$. The transformation to the embedding basis indices $i,j$ is performed as:
\begin{equation}
L_{ij}^{\mathbf{k}_J} = \frac{1}{{N_{\mathbf{k}}}} \sum_{\mathbf{k}_p, \mathbf{k}_q} (C_{\mu i}^{\mathbf{k}_p})^\dagger \, L_{\mu\nu}^{\mathbf{k}_J} \, C_{\nu j}^{\mathbf{k}_q}
\end{equation}
where $C$ are the transformation coefficients from $k$-AO to R-EO, and momentum conservation is enforced such that $\mathbf{k}_p - \mathbf{k}_q$ corresponds to the momentum of the Cholesky vector $\mathbf{k}_J$.

The resulting vectors $L_{ij}^{\mathbf{k}_J}$ are generally complex. To recover real-valued integrals suitable for the embedding solvers, we decouple the real and imaginary components and concatenate them along the auxiliary dimension: 
\begin{equation}
\tilde{L}^{\mathbf{k}_L}_{ij} = \mathrm{Re}(L^{\mathbf{k}_L}_{ij}) \oplus \mathrm{Im}(L^{\mathbf{k}_L}_{ij})
\end{equation}
By exploiting time-reversal symmetry, redundant components are eliminated, reducing the computational cost of this transformation by a factor of approximately four.

Finally, the projection of the global auxiliary basis onto the spatially compact fragment subspace introduces significant linear dependence, making the effective auxiliary dimension unnecessarily large. We compress this dimension by constructing the metric matrix of the Cholesky vectors in the embedding space:
\begin{equation}
M_{JJ'} = \sum_{ij} \tilde{L}_{ij}^J \tilde{L}_{ij}^{J'}
\end{equation}
Diagonalization of $M$ yields a set of eigenvectors $U$. We retain only those eigenvectors with eigenvalues exceeding a numerical threshold, yielding a compact set of reduced Cholesky vectors $l_{ij}^\alpha$:
\begin{equation}
l_{ij}^\alpha = \sum_{J} U_{J\alpha} \tilde{L}_{ij}^J
\end{equation}
This entire construction procedure is fully parallelized using Message Passing Interface~(MPI) and accelerated on GPUs, ensuring high efficiency even for large supercells.

\subsection*{High-level Embedding Solvers}

The embedding Hamiltonian is solved using a combination of the RPA~\cite{renRandomphaseApproximation2012,RPA} and phaseless Auxiliary-Field Quantum Monte Carlo (AFQMC)~\cite{doi:10.1021/acs.jctc.2c00802, maloneIpiePythonBased2023,jiangImprovedModularity2024, huang2024gpuacceleratedauxiliaryfieldquantummonte,mottaInitioComputationsMolecular2018,zhangQuantumMonteCarlo2003a}. For long-range correlation, we employ a density-fitted RPA within the Adiabatic Connection Fluctuation Dissipation Theorem (ACFDT) formalism~\cite{annurev:/content/journals/10.1146/annurev-physchem-040215-112308}~(see Section~\ref{SI-subsec:rpa}). Crucially, thermal smearing is explicitly incorporated via the occupation factor $f_{ia}$ in the polarizability response kernel:

\begin{equation}\label{eq:chi0}
\chi_{0,ia}(\omega) = \frac{2 f_{ia} (\epsilon_i - \epsilon_a)}{\omega^2 + (\epsilon_i - \epsilon_a)^2},
\end{equation}

where $i$ and $a$ denote occupied and virtual orbitals with energies $\epsilon_i$ and $\epsilon_a$, and occupations $f_i, f_a \in [0,1]$. The term $f_{ia} = f_i - f_a$ accounts for the fractional occupations arising from thermal smearing.

Strong local correlations are treated using phaseless AFQMC, utilizing a thermally smeared single-Slater-determinant as the trial wavefunction. The standard one-body Green’s function~\cite{doi:10.1021/acs.jctc.2c00802} for a walker $z$, $G_{pq}^{z}$, is generalized to support fractional occupations:

\begin{equation}\label{eq:smear_green}
G_{pq}^{z} = \big( \mathbf{C}_{\phi_z} (\mathbf{C}_{\psi_o}^{\dagger}\mathbf{C}_{\phi_z})^{-1} 
\mathbf{f}_{\psi_o} \mathbf{C}_{\psi_o}^{\dagger} \big)_{qp},
\end{equation}

where $p$ and $q$ label embedding basis functions, $\mathbf{C}_{\phi_z}$ contains the coefficients of the walker states $\phi_z$ in the chosen one-particle basis, $\mathbf{C}_{\psi_o}$ denotes coefficients of the occupied (or partially occupied) orbitals $\psi_o$, and $\mathbf{f}_{\psi_o}$ is a diagonal matrix of their occupations~($0 \le f_p \le 1$). The corresponding correlation energy is evaluated as 

\begin{equation}\label{eq:ecorr_afqmc}
\begin{aligned}
E_{\mathrm{corr}}^{z} ={} &
\sum_{i,j,a,b} \big(G_{ia}^{z}-G^{0}_{ia}\big)\big(G_{jb}^{z}-G^{0}_{jb}\big)
\big[2(ia|jb)-(ib|ja)\big] \\
&+ \sum_{i,a} \big(G_{ia}^{z}-G^{0}_{ia}\big)\,F^{0}_{ia},
\end{aligned}
\end{equation}

where $i,j$ denote occupied (or partially occupied) orbitals, $a,b$ denote virtual (or partially virtual) orbitals, and $z$ is the walker index, $G^{0}_{ia}$ and $F^{0}_{ia}$ represent the mean-field Green’s function and Fock matrix elements, respectively. The second term in Eq.~\eqref{eq:ecorr_afqmc} represents a correction arising solely from the non-idempotency of the density matrix due to fractional occupations and it vanishes for gapped systems. The detailed discussion on the smearing-adapted AFQMC is given in the Section~\ref{SI-subsec:afqmc} of the Supplementary Information.

\subsection*{Fragment Correlation and Energy Formulation}

To rigorously extract fragment correlation energies in the presence of fractional occupations, we employ a block-diagonal projector $\mathbf{P}^F$ that acts differently on fully and partially occupied states (see Section~\ref{SI-subsec:solvers} of the Supplementary Information). Within this scheme, fully occupied orbitals are allowed to mix only within their subspace, while each partially occupied orbital is projected independently onto its fragment counterpart. This construction effectively eliminates spurious off-diagonal contributions, ensuring a well-defined fragment energy estimator that is applied consistently across both RPA and AFQMC solvers. In AFQMC, the projected Green’s function for a walker $z$ is given by

\begin{equation}\label{eq:gf_proj}
G_{\tilde{i}a}^{z} = \sum_{i} (\mathbf{P}^F_{f})_{\tilde{i}i}\, G_{ia}^{z},
\end{equation}
where $f$ denotes the fragment index and $\tilde{i}$ represents the fragment-projected counterpart of orbital $i$. This projection ensures that local correlation energies are evaluated strictly within the fragment space. An analogous projection is applied in RPA, providing a unified estimator of fragment correlation energies under thermal smearing.

With fragment energies defined consistently for both solvers, the total energy is obtained by combining the global RPA baseline with fragment-wise AFQMC corrections, 
\begin{equation}
    \begin{split}
        E_{\mathrm{total}} &= E^{\mathrm{global}}_{\mathrm{RPA}} + \sum_{f} \delta E_f \\
       &= E^{\mathrm{global}}_{\mathrm{RPA}} + \sum_{f} \left( E_{f}^{\mathrm{AFQMC}} - E_{f}^{\mathrm{RPA}} \right),
    \end{split}
\end{equation} 
where $E^{\mathrm{global}}_{\mathrm{RPA}}$ is the global RPA energy of the full periodic system, $\delta E_{f}$ represents the local many-body correction (AFQMC relative to RPA) for fragment $f$. We note the global $k$-RPA is accelerated via a fully distributed GPU algorithm with MPI parallelization, enabling the treatment of systems containing tens of thousands of orbitals.

\section*{Author Contributions} 

The manuscript was written through contributions from all authors. All authors have reviewed and approved the final version. H. Pham conceptualized the embedding framework, implemented the corecodebase, and performed numerical simulations and analysis. C. Cao proposed the project, developed the framework, and performed numerical simulations and analysis. Z. Guo and Z. Huang supported code development and GPU optimization. Y. Zhang worked on the GPU-accelerated k-dapted RPA implementation. X. Wen and Z. Huang provided insightful discussions and suggestions during the project. J. Chen contributed valuable discussions and insights. D. Lv conceptualized the project, provided key discussions, managed the project, and provided overall supervision. \\

\section*{Acknowledgements}
The authors acknowledge Dr. Hang Li and ByteDance Seed AI for Science teams for their invaluable support. The authors thank Dr. Hongzhou Ye, Dr. Hai Xiao, Dr. Han-Shi Hu, Dr. Xinguo Ren, Dr. Peifeng Liu, Dr. Qing Zhao and Dr. Guo P. Chen for insightful scientific discussions throughout this project. We also thank Yunze Qiu, Junjie Yang, Jia Gao, Yibo Wu and Zechuan Liu for their valuable feedback and suggestions. Ji Chen is supported by the National Key R\&D Program of China~(2021YFA1400500) and National Science Foundation of China~(12334003).

\bibliography{refs}


\end{document}


\maketitle 
\tableofcontents
\newpage

\section{Quantum chemical methods with thermal smearing}

\subsection{Notation}

\begin{center}
    \begin{tabular}{l l}
        \hline
        \textbf{Notation} & \textbf{Description} \\
        \hline
        $p$, $q$, $r$, $s$, etc. & indices for atomic orbitals \\
        $i$, $j$, $k$, $l$, etc. & indices for occupied orbitals \\
        $a$, $b$, $c$, $d$, etc. & indices for virtual orbitals \\
        $\alpha$, $\beta$, $\gamma$, etc. & indices for auxiliary vectors \\
        $x$, $y$, $z$, etc. & indices for walkers \\
        $N_{ao}$  & the number of atomic orbitals or basis functions\\
        $N_{eo}$  & the number of embedding orbitals\\
        $N_e$ & the number of electrons \\
        $N_{\text{aux}}$ & the number of the auxiliary basis \\
        $N_{\text{walkers}}$  & the number of walkers \\
        $N_{\text{dets}}$ & the number of determinants \\
        \hline
    \end{tabular}
\end{center}

\subsection{Periodic quantum chemistry}\label{subsec:periodic_qc}

Many-body simulations of periodic solids within FEMION are formulated in reciprocal space, where translational symmetry enables an efficient and systematically improvable treatment of electron correlation:

\begin{equation}\label{kHam1}
\hat{H} = \sum_{p,q} \sum_{\textbf{k}_p, \textbf{k}_q} h^{\textbf{k}_p \textbf{k}_q}_{pq} \hat{a}^{\textbf{k}_p \dagger}_p \hat{a}^{\textbf{k}_q}_q 
+ \frac{1}{2} \sum_{p,q,r,s} \sum_{\textbf{k}_p, \textbf{k}_q, \textbf{k}_r, \textbf{k}_s} g^{\textbf{k}_p \textbf{k}_q \textbf{k}_r \textbf{k}_s}_{pqrs} \hat{a}^{\textbf{k}_p \dagger}_p \hat{a}^{\textbf{k}_r \dagger}_r \hat{a}^{\textbf{k}_s}_s \hat{a}^{\textbf{k}_q}_q
\end{equation}

where $h^{\textbf{k}_p \textbf{k}_q}_{pq}$ and $g^{\textbf{k}_p \textbf{k}_q \textbf{k}_r \textbf{k}_s}_{pqrs}$ denote the one- and two-electron integrals in momentum space, respectively. Here, $\hat{a}^{\textbf{k}_p\dagger}_p$ ($\hat{a}^{\textbf{k}_q}_q$) is the fermionic creation (annihilation) operator for the $p$-th orbital in $\textbf{k}_p$ ($\textbf{k}_q$), sampled in momentum space (\textbf{k}-space). The indices $p$, $q$, $r$, and $s$ label crystalline orbitals. Due to translational symmetry, $\textbf{k}_p - \textbf{k}_q = 0$ for one-body terms and $\textbf{k}_p + \textbf{k}_r - \textbf{k}_s - \textbf{k}_q = 0$ for two-body terms, reflecting the conservation of crystal momentum.

For periodic systems, atomic Bloch orbitals, or \textbf{k}-adapted atomic orbitals (AOs), are defined as the Fourier transform of contracted Gaussian-type orbitals (cGTOs):
\begin{equation}
\phi^{\textbf{k}}_{\mu}(\textbf{r})=\sum_{\textbf{R}} e^{\rm{i}\textbf{k}\cdot\textbf{R}} \, \phi_{\mu}(\textbf{r}-\textbf{R})
\label{kBasis}
\end{equation}
where $\mu$ denotes the AO index within the unit cell, $\textbf{R}$ is a real-space lattice vector, and $\textbf{k}$ is a crystal momentum. Crystalline orbitals are then obtained as linear combinations of atomic Bloch orbitals,
\begin{equation}
\psi^{\textbf{k}}_p(\textbf{r}) = \sum_{\mu} C^{\textbf{k}}_{\mu p} \, \phi^{\textbf{k}}_{\mu}(\textbf{r})
\end{equation}
where $C^{\textbf{k}}_{\mu p}$ are the crystalline orbital coefficients.

\subsection{Mean-field}\label{subsec:meanfield}
In mean-field calculations, the one-particle density matrix (1-RDM) is a central quantity that describes the probability of finding an electron in a given orbital. In the thermal smearing approach, rather than assigning orbitals binary occupation numbers (0 or 2), fractional occupations are used to smooth the electron occupancy near the Fermi level. This is particularly useful for metallic or near-metallic systems.

The 1-RDM is typically evaluated as
\begin{equation}\label{eq:1RDM}
\gamma_{ij} = \sum_{\mu} f_{\mu} \, C_{i\mu} \, C_{j\mu}^*
\end{equation}
where $C_{i\mu}$ are the coefficients of the molecular orbitals in the chosen basis, and $f_\mu$ denotes the occupation number of orbital $\mu$ as determined by the smearing function. The fractional occupations $f_\mu$ enable a smoother description of the electron density, which can improve convergence and provide a more realistic treatment of systems with partially occupied states.

In periodic solid-state calculations, the electronic structure is evaluated over a discrete grid of k-points in the Brillouin zone. In this context, the 1-RDM is constructed by summing the contributions from all k-points, with each contribution weighted by a fractional occupation obtained from a smearing scheme (in this work, Gaussian smearing). This approach smooths the electronic occupancy near the Fermi level, which is crucial for accurately treating metals and gapless systems. For a periodic system, the 1-RDM is given by
\begin{equation}\label{eq:1RDM_periodic}
\gamma_{ij} = \frac{1}{N_k} \sum_{\mathbf{k}} \sum_{n} f_{n\mathbf{k}} \, C_{i,n}(\mathbf{k}) \, C_{j,n}^*(\mathbf{k})
\end{equation}
where $N_k$ is the number of k-points, $f_{n\mathbf{k}}$ is the fractional occupation for band $n$ at k-point $\mathbf{k}$, and $C_{i,n}(\mathbf{k})$ are the Bloch orbital coefficients. This formulation ensures accurate integration over the Brillouin zone, yielding a robust description of the electron density in periodic systems.

This reciprocal-space formulation provides the foundation for the thermal smearing, RPA correlation treatment, and fragment embedding strategies introduced in the following sections.

\subsection{Direct Random-phase approximation~(dRPA)}\label{subsec:rpa}

\subsubsection{Formulation}
Within the adiabatic-connection fluctuation-dissipation theorem (ACFDT), the random-phase approximation (RPA)~\cite{RPA, RPA_tianyu,renRandomphaseApproximation2012} ground-state energy, $E_{\mathrm{RPA}}$, is written as the sum of the correlation energy, $E_{c}$, and the Hartree–Fock (HF) energy, $E_{\mathrm{EXX}}$:
\begin{equation}\label{eq:ACFDT_RPA}
E_{\mathrm{RPA}} = E_{c} + E_{\mathrm{EXX}}
\end{equation}

Here, $E_{\mathrm{EXX}}$ is evaluated non-self-consistently using Kohn–Sham orbitals from density-functional theory.

The ACFDT-RPA correlation energy is given by
\begin{equation}\label{eq:dRPAcorrelation}
E_c^{\text{dRPA}} = \frac{1}{2\pi}\int_0^{\infty} d\omega\, \mathrm{Tr}\Big[\ln\Big(1-\chi_0(i\omega)v\Big) + \chi_0(i\omega)v\Big]
\end{equation}
where $\chi_0(i\omega)$ is the non-interacting density–density response function (Lindhard function), $v(\mathbf{r},\mathbf{r}') = \frac{1}{|\mathbf{r}-\mathbf{r}'|}$ is the Coulomb interaction kernel, and $\mathrm{Tr}$ denotes the trace over the spatial coordinates $\mathbf{r}$ and $\mathbf{r}'$.

Within density fitting (DF), using electron repulsion integrals (ERIs) expressed with $v_P^{ia}$, the dRPA correlation energy can be written as
\begin{equation}\label{eq:dRPAcorrelation_DF}
E_c^{\text{dRPA}} = \frac{1}{2\pi}\int_0^{\infty} d\omega\, \mathrm{Tr}\Big[\ln\big(\mathbf{1}-\mathbf{\Pi}(\omega)\big) + \mathbf{\Pi}(\omega)\Big]
\end{equation}

As long as $\mathbf{1}-\mathbf{\Pi}(\omega)$ is semipositive definite, the following relation holds:
\begin{equation}\label{eq:trace_relation}
\mathrm{Tr}\Big[\ln\big(\mathbf{1}-\mathbf{\Pi}(\omega)\big) + \mathbf{\Pi}(\omega)\Big] 
= \ln\Big[\det\big(\mathbf{1}-\mathbf{\Pi}(\omega)\big)\Big] + \mathrm{Tr}\big(\mathbf{\Pi}(\omega)\big)
\end{equation}

Thus, the dRPA correlation energy can be expressed as
\begin{equation}\label{eq:dRPAcorrelation_DF_det}
E_c^{\text{dRPA}} = \frac{1}{2\pi}\int_0^{\infty} d\omega\, 
\Big[\ln\Big(\det\big(\mathbf{1}-\mathbf{\Pi}(\omega)\big)\Big) + \mathrm{Tr}\big(\mathbf{\Pi}(\omega)\big)\Big]
\end{equation}
which is both more efficient and numerically stable.

For a restricted mean-field reference, the dielectric matrix $\mathbf{\Pi}$ in spatial orbitals is given by
\begin{equation}\label{eq:dielectric_matrix}
\Pi_{PQ}(\omega) = 2\sum_{ia} v_P^{ia}\,\chi_{0,ia}(\omega)\,v_Q^{ia*}
\end{equation}
where the density response kernel is defined as
\begin{equation}\label{eq:density_response}
\chi_{0,ia}(\omega) = \frac{2(\epsilon_i - \epsilon_a)}{\omega^2 + (\epsilon_i - \epsilon_a)^2}
\end{equation}

\subsubsection{Thermal smearing density response kernel}

If a smearing function is employed in the restricted mean-field, a factor 
\begin{equation}\label{eq:f_ia}
f_{ia} = f_i - f_a
\end{equation}
is introduced into the density response kernel, modifying it as follows:
\begin{equation}\label{eq:chi0}
\chi_{0,ia}(\omega) = \frac{2\, f_{ia}\, (\epsilon_i - \epsilon_a)}{\omega^2 + (\epsilon_i - \epsilon_a)^2}
\end{equation}

Note that the indices $i$ and $a$ are not limited to occupied and virtual orbitals; rather, they run over all possible orbitals. The physical interpretation of this modification remains nontrivial, especially since $\Pi_{PQ}(\omega)$ also includes intraband (same-band) transitions.

Furthermore, $f_{ia}$ is sparse because the blocks corresponding to fully occupied–occupied ($o\text{-}o$) and fully virtual–virtual ($v\text{-}v$) orbital pairs vanish. Additionally, $f_{ia}$ is antisymmetric, i.e.,
\begin{equation}
f_{ia} = - f_{ai}
\end{equation}
These properties are exploited in the implementation to reduce computational cost.

For periodic solids, this formulation must further account for Brillouin zone integration. In this context, the expression for the dielectric matrix becomes
\begin{equation}\label{eq:dielectric_matrix_periodic}
\Pi_{PQ}(\omega, \mathbf{k}) = \frac{2}{N_k} \sum_{n,m} v_{P}^{n\mathbf{k},m\mathbf{k}}\, \chi_{0,nm}(\mathbf{k};\omega)\, \Big(v_{Q}^{n\mathbf{k},m\mathbf{k}}\Big)^*
\end{equation}
where $N_k$ is the number of k-points, and the indices $n$ and $m$ refer to electronic bands. The density response kernel is now defined as
\begin{equation}\label{eq:chi0_periodic}
\chi_{0,nm}(\mathbf{k};\omega) = \frac{2\, f_{nm}(\mathbf{k}) \, \big(\epsilon_{n\mathbf{k}} - \epsilon_{m\mathbf{k}}\big)}{\omega^2 + \big(\epsilon_{n\mathbf{k}} - \epsilon_{m\mathbf{k}}\big)^2}
\end{equation}
with
\begin{equation}\label{eq:f_nm}
f_{nm}(\mathbf{k}) = f_{n\mathbf{k}} - f_{m\mathbf{k}}
\end{equation}
where $f_{n\mathbf{k}}$ denotes the fractional occupation of band $n$ at k-point $\mathbf{k}$. In this formulation, $v_{P}^{n\mathbf{k},m\mathbf{k}}$ represents the density-fitted electron repulsion integrals in the periodic setting. This extended formulation ensures accurate Brillouin zone integration and captures the $\mathbf{k}$-dependent nature of electronic excitations in periodic solids.

\subsection{Phaseless auxiliary-field quantum Monte Carlo (ph-AFQMC)}\label{subsec:afqmc}
\subsubsection{AFQMC with singlet Slater-determinant (SD) trial}

Here, the trial wavefunction $\Psi_T$ is assumed to be a single Slater determinant (represented by $\mathbf{C}_{\psi_0}$); extensions to multi-determinant trial states will be considered in future work.

For the two-electron integrals, a Cholesky decomposition is employed:
\begin{equation}\label{eq:chol}
(pr|qs) = \sum_{\alpha=1}^{N_{\text{aux}}} L_{pr}^{\alpha} \, L_{qs}^{\alpha}
\end{equation}
It is often convenient to form half-rotated Cholesky vectors by rotating one of the atomic orbital indices into the occupied molecular orbital space:
\begin{equation}\label{eq:half_rotated}
L_{ir}^{\alpha} = \sum_p (\mathbf{C}_{\psi_0}^{\dagger})_{ip} \, L_{pr}^{\alpha}
\end{equation}

The overlap between the trial wavefunction $\Psi_T$ and a walker $\phi_z$ is given by
\begin{equation}\label{eq:overlap}
\langle \Psi_T|\phi_z \rangle = \det\Bigl(\mathbf{C}_{\psi_0}^{\dagger}\mathbf{C}_{\phi_z}\Bigr)
= \det\Bigl(\mathbf{C}_{\phi_z}^{T}\mathbf{C}_{\psi_0}^{*}\Bigr)
\end{equation}

The one-body Green’s function for walker $z$  is defined as
\begin{equation}\label{eq:green}
G_{pq}^{z} = \frac{\langle \Psi_T|\hat{a}_p^{\dagger}\hat{a}_q|\phi_z \rangle}{\langle \Psi_T|\phi_z \rangle}
\end{equation}
which can be expressed as
\begin{equation}\label{eq:green_expanded}
G_{pq}^{z} = \Bigl(\mathbf{C}_{\phi_z}\, (\mathbf{C}_{\psi_0}^{\dagger}\mathbf{C}_{\phi_z})^{-1}\, \mathbf{C}_{\psi_0}^{\dagger}\Bigr)_{qp}
= \sum_{j=1}^{N_{\text{occ}}} \theta_{jp}^{z} \, (\mathbf{C}_{\psi_0}^{\dagger})_{jq}
\end{equation}
with the walker-dependent part of the Green’s function defined as
\begin{equation}\label{eq:theta}
\theta_{jp}^{z} = \Bigl(\mathbf{C}_{\phi_z}\, (\mathbf{C}_{\psi_0}^{\dagger}\mathbf{C}_{\phi_z})^{-1}\Bigr)_{pj}
\end{equation}

The corresponding optimal force-bias potential is then given by
\begin{equation}\label{eq:force_bias}
\bar{x}_{\alpha}^z = \frac{\langle \Psi_T|\hat{a}_p^{\dagger}\hat{a}_q|\phi_z \rangle}{\langle \Psi_T|\phi_z \rangle}\, L_{pr}^{\alpha}
= G_{pq}^{z}\, L_{pq}^{\alpha}
= \sum_{pj} \theta_{jp}^{z}\, L_{pj}^{\alpha}
\end{equation}

The local energy for walker $z$ is decomposed into one-body and two-body contributions:
\begin{equation}\label{eq:local_energy}
E^z_{L} = E^z_{1,L} + E^z_{2,L}
\end{equation}
The one-body part is
\begin{equation}\label{eq:local_energy_1body}
E^z_{1,L} = \sum_{\alpha}\sum_{ip} L_{ip}^{\alpha}\, \theta_{ip}^{z}.
\end{equation}
The two-body part consists of a classical Coulomb term $E^z_{L,J}$ and an exchange term $E^z_{L,K}$:
\begin{equation}\label{eq:local_energy_2body}
E^z_{2,L} = E^z_{L,J} + E^z_{L,K}
\end{equation}
with
\begin{equation}\label{eq:coulomb}
E^z_{L,J} = \sum_{\alpha}\sum_{ij}\sum_{pq} L_{ip}^{\alpha}\, L_{jq}^{\alpha}\, \theta_{ip}^{z}\, \theta_{jq}^{z}
= \sum_{\alpha}\left(\sum_{ip} L_{ip}^{\alpha}\, \theta_{ip}^{z}\right)
\left(\sum_{jq} L_{jq}^{\alpha}\, \theta_{jq}^{z}\right)
\end{equation}
and
\begin{equation}\label{eq:exchange}
E^z_{L,K} = -\frac{1}{2}\sum_{\alpha}\sum_{ij}\sum_{pq} L_{ip}^{\alpha}\, L_{jq}^{\alpha}\, \theta_{iq}^{z}\, \theta_{jp}^{z}
= -\frac{1}{2}\sum_{\alpha}\sum_{ij}\left(\sum_{p} L_{ip}^{\alpha}\, \theta_{jp}^{z}\right)
\left(\sum_{q} L_{jq}^{\alpha}\, \theta_{iq}^{z}\right)
\end{equation}

Finally, the total energy in the AFQMC framework is obtained as a weighted average over the walkers:
\begin{equation}\label{eq:total_energy}
E = \frac{\sum_{z} \omega_z \, E^z_{L}}{\sum_{z} \omega_z}
\end{equation}
This formulation provides the reference ph-AFQMC framework used throughout this work, upon which thermal smearing and embedding extensions are built.

\subsubsection{Thermal smearing SD trial}

In the case of smearing occupancies, the 1-RDM is given by

\begin{equation}\label{eq:1RDM_smear}
D_{pq} = \mathbf{C}_{\psi_0}\, \mathbf{f}_{\psi_0}\, \mathbf{C}_{\psi_0}^{\dagger}
\end{equation}
where $\mathbf{f}_{\psi_0}$ is the diagonal occupancy matrix for the smeared trial wavefunction.

Assuming that the same $\mathbf{f}_{\psi_0}$ is used for both the trial and the Walker states, the Green function of one body can be computed as

\begin{equation}\label{eq:green_smear}
G_{pq}^{z} = \Bigl(\mathbf{C}_{\phi_z}\, (\mathbf{C}_{\psi_0}^{\dagger}\mathbf{C}_{\phi_z})^{-1}\, \mathbf{f}_{\psi_0}\, \mathbf{C}_{\psi_0}^{\dagger}\Bigr)_{qp}
\end{equation}
\clearpage

\section{Fragment Embedding for Metals and Insulators with On-site and Nonlocal Correlation}

\subsection{Domain-localized correlated bath space}\label{subsec:domain}
The bath natural orbital~(BNO) method for extending the density matrix embedding theory~(DMET) bath can become computationally demanding as the number of k-points increases, particularly for metallic systems. To mitigate this cost while retaining systematic improvability, we employ a Boughton–Pulay (BP) domain to restrict the subspaces used for the bath extension. The BP domain is a local orbital selection scheme, originally proposed by Boughton and Pulay~\cite{BPdomain}, which identifies the most relevant atomic basis functions contributing to a fragment by applying a density-based threshold. 
This allows the orbital space to be truncated systematically while preserving the dominant contributions to the fragment density.

We first construct a DMET cluster using the conventional algorithm based on the mean-field 1-RDM, which decomposes the full orbital space into
\begin{equation}\label{eq:orbital_decomposition}
[\mathbf{C}_{frag}\;|\;\mathbf{C}_{bath}\;|\;\mathbf{C}_{env}^{occ}\;|\;\mathbf{C}_{env}^{vir}]
\end{equation}
where $\mathbf{C}_{frag}$ are fragment orbitals, $\mathbf{C}_{bath}$ are bath orbitals, $\mathbf{C}_{env}^{occ}$ are environment occupied orbitals, and $\mathbf{C}_{env}^{vir}$ are environment virtual orbitals.  

Next, a BP domain for the fragment is constructed using a threshold $T_{BP}$, following Kállay’s work. Intuitively, the BP domain for the fragment is defined as the set of atoms (or $N_{BP}$ atomic basis functions) that together account for up to $100 \times T_{BP}\%$ of the fragment density. As $T_{BP} \rightarrow 1$, the BP domain approaches the full system (supercell), recovering the full BNO limit. In our work, $T_{BP}=0.95$ is used as the default, and this parameter serves as the main tuning parameter for the size of embedding orbitals.

A projector onto this domain, $\hat{P}_{BP}$, is then constructed. In a localized orbital basis, this projector is given by
\begin{equation}\label{eq:bp_projector}
\mathbf{P}_{BP} = \mathbf{C}_{ao,lo}^\dagger\,\mathbf{S}_{BP,ao}^\dagger\,\mathbf{S}_{BP}^{-1}\,\mathbf{S}_{BP,ao}\,\mathbf{C}_{ao,lo}
\end{equation}
where $\mathbf{C}_{ao,lo}$ transforms from atomic orbital (AO) to localized orbital (LO) basis, $\mathbf{S}_{BP,ao}$ is the AO overlap matrix restricted to the BP domain, and $\mathbf{S}_{BP}$ is the overlap matrix within the BP domain.

Using $\hat{P}_{BP}$, we can divide the environment-occupied orbitals $\mathbf{C}_{env}^{occ}$ into two sets:
\begin{equation}\label{eq:env_occ_split}
\mathbf{C}_{env}^{occ} = \Bigl[\mathbf{C}_{env}^{'occ} \;\Big|\; \mathbf{C}_{env}^{''occ}\Bigr]
\end{equation}
Here, $\mathbf{C}_{env}^{'occ}$ comprises those orbitals that strongly overlap with the BP domain. This is achieved by first projecting $\mathbf{C}_{env}^{occ}$ onto the BP domain, then computing the overlap matrix
\begin{equation}\label{eq:overlap_occ}
\mathbf{S}_{BP}^{occ} = \mathbf{C}_{env}^{occ\,\dagger}\,\mathbf{P}_{BP}\,\mathbf{C}_{env}^{occ}
\end{equation}
and diagonalizing it:
\begin{equation}\label{eq:diag_occ}
\mathbf{S}_{BP}^{occ} = \mathbf{U}^{occ}\,\sigma\, \mathbf{U}^{occ\,\dagger}
\end{equation}
The BP-localized occupied set is then defined as
\begin{equation}\label{eq:BP_occ}
\mathbf{C}_{env}^{'occ} = \mathbf{C}_{env}^{occ}\,\mathbf{U}^{occ}_{\lvert\sigma\rvert > \text{thr}}
\end{equation}
with the remainder given by
\begin{equation}\label{eq:BP_occ_rem}
\mathbf{C}_{env}^{''occ} = \mathbf{C}_{env}^{occ}\,\mathbf{U}^{occ}_{\lvert\sigma\rvert < \text{thr}}
\end{equation}
Similarly, the BP-localized virtual set is obtained by forming
\begin{equation}\label{eq:overlap_vir}
\mathbf{S}_{BP}^{vir} = \mathbf{C}_{env}^{vir\,\dagger}\,\mathbf{P}_{BP}\,\mathbf{C}_{env}^{vir}
\end{equation}
diagonalizing
\begin{equation}\label{eq:diag_vir}
\mathbf{S}_{BP}^{vir} = \mathbf{U}^{vir}\,\sigma\, \mathbf{U}^{vir\,\dagger}
\end{equation}
and defining
\begin{equation}\label{eq:BP_vir}
\mathbf{C}_{env}^{'vir} = \mathbf{C}_{env}^{vir}\,\mathbf{U}^{vir}_{\lvert\sigma\rvert > \text{thr}}
\end{equation}
\begin{equation}\label{eq:BP_vir_rem}
\mathbf{C}_{env}^{''vir} = \mathbf{C}_{env}^{vir}\,\mathbf{U}^{vir}_{\lvert\sigma\rvert < \text{thr}}
\end{equation}

At this stage, an embedding space can be constructed as
\begin{equation}\label{eq:embedding_space}
[\mathbf{C}_{frag}\;|\;\mathbf{C}_{bath}\;|\;\mathbf{C}_{env}^{'occ}\;|\;\mathbf{C}_{env}^{'vir}]
\end{equation}
which has a maximum size of $2\,N_{frag}+2\,N_{BP}$ and is defined as the extended DMET cluster.

This extended cluster can be further refined or truncated using an approximate MP2 calculation. Similar to the BNO approach, the DMET orbitals are split into
\begin{equation}\label{eq:DMET_split}
[\mathbf{C}_{frag}\;|\;\mathbf{C}_{bath}] = [\mathbf{C}_{DMET}^{occ}\;|\;\mathbf{C}_{DMET}^{vir}]
\end{equation}
by projecting the 1-RDM into the DMET space and diagonalizing to obtain natural orbitals and their occupancies. In the case of a smeared mean-field, where the occupancies are fractional, the Aufbau principle is used instead of natural orbital occupancies to split the DMET space into occupied and virtual parts. The extended DMET space then becomes
\begin{equation}\label{eq:extended_DMET}
[\mathbf{C}_{DMET}^{occ}\;|\;\mathbf{C}_{DMET}^{vir}\;|\;\mathbf{C}_{env}^{'occ}\;|\;\mathbf{C}_{env}^{'vir}]
\end{equation}
which is partitioned into two subspaces:
\begin{equation}\label{eq:DMET_partition}
[\mathbf{C}_{env}^{'occ}\;|\;\mathbf{C}_{DMET}^{vir}] \quad \text{and} \quad [\mathbf{C}_{env}^{'vir}\;|\;\mathbf{C}_{DMET}^{occ}]
\end{equation}
An approximate MP2 amplitude for the subspace $[\mathbf{C}_{env}^{'occ}\;|\;\mathbf{C}_{DMET}^{vir}]$ is computed as
\begin{equation}\label{eq:MP2_amplitude}
t_{ij}^{ab} = \frac{\sum (ia|L)(L|jb)}{\epsilon_i + \epsilon_j - \epsilon_a - \epsilon_b}
\end{equation}
where $\epsilon_i$ and $\epsilon_j$ are the orbital energies of $\mathbf{C}_{env}^{'occ}$, and $\epsilon_a$ and $\epsilon_b$ are the orbital energies of $\mathbf{C}_{DMET}^{vir}$. These energies are obtained by diagonalizing the Fock matrix within their respective subspaces (quasi-canonicalization).

For periodic systems, the integrals $(L|ia)$ are expressed as
\begin{equation}\label{eq:integrals_periodic}
(L|ia) = \Bigl[(L_{0}|ia)\;\Bigl|\;(L_{1}|ia)\;\Bigl|\;\dots\;\Bigl|\;(L_{k_L}|ia)\Bigr]
\end{equation}
and the MP2 amplitude is given by
\begin{equation}\label{eq:MP2_periodic}
t_{ij}^{ab}(kL) = -\frac{\sum (ia|L_{k_L})(L_{k_L}|jb)}{\epsilon_a + \epsilon_b - \epsilon_i - \epsilon_j}
\end{equation}
\begin{equation}\label{eq:MP2_total}
t_{ij}^{ab} = \sum_{k_L} t_{ij}^{ab}(kL)
\end{equation}
This algorithm ensures that the full set of $(L|ia)$ integrals is never stored, since the $L$ dimension scales linearly with the number of k-points.

Finally, the $[\text{occ}|\text{occ}]$ block of the 1-RDM is computed using the MP2 amplitudes $t_{ij}^{ab}$. Following the procedure in the BNO method, the set $\mathbf{C}_{env}^{'occ}$ is split into two subsets based on the natural occupancies obtained by diagonalizing the occ-occ block of the 1-RDM:
\begin{equation}\label{eq:split_occ}
\mathbf{C}_{env}^{'occ} = [\mathbf{C}_{env}^{'occ,1}\;|\;\mathbf{C}_{env}^{'occ,2}]
\end{equation}
and similarly, the $[\text{vir}|\text{vir}]$ block for the subspace $[\mathbf{C}_{env}^{'vir}\;|\;\mathbf{C}_{DMET}^{occ}]$ splits $\mathbf{C}_{env}^{'vir}$ into
\begin{equation}\label{eq:split_vir}
\mathbf{C}_{env}^{'vir} = [\mathbf{C}_{env}^{'vir,1}\;|\;\mathbf{C}_{env}^{'vir,2}]
\end{equation}
The final embedding orbital space is then assembled as
\begin{equation}\label{eq:final_embedding}
[\mathbf{C}_{frag}\;|\;\mathbf{C}_{bath}\;|\;\mathbf{C}_{env}^{'occ,1}\;|\;\mathbf{C}_{env}^{'vir,1}]
\end{equation}
When $T_{BP} \rightarrow 1$, this procedure approaches the BNO limit for a given bath threshold. In practice, all eigenvalue thresholds are fixed at $1\times10^{-9}$, so that users need only tune $T_{BP}$ to control the embedding size and, consequently, the computational resources required.

\subsection{Embedding Hamiltonian}\label{subsec:embham}

The one-electron integrals for the embedded subspace are computed as
\begin{equation}
\tilde{h}_{ij} = C_{pi}^{\mathbf{k}\dagger}\,(h^{\mathbf{k}}_{pq} + V^{\mathbf{k},\mathrm{eff}}_{pq})\,C^{\mathbf{k}}_{qj}
- \sum_{kl}^{\mathrm{emb}} D_{kl}^{\mathrm{emb}}\left[(ij|lk) - \tfrac{1}{2}(ik|lj)\right]
\end{equation}
where $C$ is the projection operator that maps the AO-basis integrals ($h_{pq}^{\text{AO}}$) to the embedding subspace, and $V^{\mathrm{eff}}$ represents the mean-field potential contribution from the DMET bath. The projection operator $C$ ensures that the subspace retains interactions between the embedded cluster and the bath, while $V^{\mathrm{eff}}$ corrects for the double-counting of Coulomb and exchange interactions. This formulation guarantees that the embedding Hamiltonian remains consistent with the mean-field environment, preserving the self-consistency of the total system.

Direct computation of the full four-index two-electron integral tensor $(ij|lk)$ is prohibitively expensive for large embeddings due to its $O(N^4)$ memory scaling. Moreover, for periodic systems, sampling in momentum space inherently generates complex-valued electron repulsion integrals, which are incompatible with conventional molecular high-level solvers designed for real integrals.

To circumvent these limitations, we transfer the Cholesky-decomposed electron repulsion integrals (cderi) directly from the k-point atomic orbital (k-AO) basis to the real-space embedding orbital (R-EO) basis. The cderi is generated during mean-field calculations, where the two-electron integrals are approximated as a sum of outer products of Cholesky vectors:
\begin{equation}
(\mu\nu|\rho\sigma) \approx \sum_L^{N_{\text{aux}}} L_{\mu\nu}^L\,L_{\rho\sigma}^L
\end{equation}
with $N_{\text{aux}}$ being the dimension of the auxiliary basis.  
The transformation is performed as
\begin{equation}
L^{\mathbf{k}_L}_{ij} = \frac{1}{N_{\mathbf{k}}} \sum_{\mathbf{k}_p,\mathbf{k}_q} C_{ip}^{\mathbf{k}_p\dagger}\,L^{\mathbf{k}_L}_{p^{\mathbf{k}_p}q^{\mathbf{k}_q}}\,C_{qj}^{\mathbf{k}_q}
\end{equation}
where $L^{\mathbf{k}_L}_{p^{\mathbf{k}_p}q^{\mathbf{k}_q}}$ are the Cholesky vectors in the k-AO basis and $N_{\mathbf{k}}$ is the number of k-points.

The resulting integrals $(L^{\mathbf{k}_L}|ij)$ in the embedding space are generally complex. To obtain real integrals, we separate the real and imaginary components and concatenate them into a single real tensor:
\begin{equation}
\tilde{L}^{\mathbf{k}_L}_{ij} = \mathrm{Re}(L^{\mathbf{k}_L}_{ij}) \oplus \mathrm{Im}(L^{\mathbf{k}_L}_{ij})
\end{equation}
Taking advantage of time-reversal symmetry, the computational cost of the transformation is reduced by a factor of approximately four.  

In this way, we obtain a compact real cderi representation for the embedding. This enables the use of efficient density-fitting high-level solvers to treat the embedding Hamiltonian.

\subsection{Impurity solvers}\label{subsec:solvers}
In this section, we describe the low- and high-level impurity solvers used to treat the embedding Hamiltonian.
\subsubsection{Low-level RPA impurity solver}
\paragraph{Cluster double excitation amplitudes}

The double excitation amplitudes $t_{ij}^{ab}$ can, in principle, be obtained by the conventional iterative procedure used in coupled cluster (CC) methods. However, there are several reasons why this approach does not apply straightforwardly to a smeared mean-field:

First, the equivalence between dRPA in its plasmon form and direct ring-CCD (drCCD) has only been proven for gapped systems. Second, in ACFDT-dRPA the density response kernel is augmented with an occupancy factor $f_{ia} = f_i - f_a$ to account for the partial contributions of orbitals to the dielectric function. This modification, while essential for a smeared mean-field, renders the derivation of cluster amplitudes via a conventional CC-like calculation nontrivial. Third, there is an additional numerical instability in solving for $t_{ij}^{ab}$ when $(\epsilon_i+\epsilon_j-\epsilon_a-\epsilon_b) \to 0$, as is typical in metallic systems.. Although this instability can be mitigated with a regularizer or level shift, it generally slows convergence and increases computational cost.

Eshuis \emph{et al.}~\cite{eshuisFastComputation2010} showed that the cluster amplitude matrix $\mathbf{T}$ can instead be computed within the ACFDT formalism using the quadrature
\begin{equation}\label{eq:T_general}
\mathbf{T} = -\frac{2}{2\pi}\int_{-\infty}^{\infty} d\omega\, \mathbf{G}(\omega)\,\mathbf{S}\,\mathbf{R}(\omega)\,\mathbf{S}^{T}\,\mathbf{G}(\omega)
\end{equation}
with
\begin{equation}\label{eq:R_general}
\mathbf{R}(\omega) = -2\,\mathbf{Q}^{-2}(\omega)\Bigl[\ln\bigl(\mathbf{1}+\mathbf{Q}(\omega)\bigr) - \mathbf{Q}(\omega)\Bigr]
\end{equation}
where $\mathbf{G}(\omega) \equiv -\chi_0(\omega)$ is the non-interacting density response kernel, $\mathbf{Q}(\omega) \equiv -\mathbf{\Pi}(\omega)$ is the negative of the dielectric matrix, and the density-fitting electron repulsion integrals (ERIs) are defined by
\[
S_P^{ia} \equiv v_P^{ia}.
\]

For a restricted mean-field, these equations simplify to
\begin{equation}\label{eq:T_restricted}
\mathbf{T} = -\frac{2}{2\pi}\int_{0}^{\infty} d\omega\, \chi_0(\omega)\,\mathbf{v}\,\mathbf{R}(\omega)\,\mathbf{v}^{T}\,\chi_0(\omega)
\end{equation}
\begin{equation}\label{eq:R_restricted}
\mathbf{R}(\omega) = -2\,\mathbf{\Pi}^{-2}(\omega)\Bigl[\ln\bigl(\mathbf{1}-\mathbf{\Pi}(\omega)\bigr) + \mathbf{\Pi}(\omega)\Bigr]
\end{equation}

For a smeared mean-field, the kernel $\chi_{0,ia}(\omega)$ is augmented with the occupancy function $f_{ia}$ as discussed previously, so that no extra factor is introduced when computing $\mathbf{T}$. This ensures consistency in the treatment of fractional occupancies between the dRPA energy and its cluster amplitude. The primary difference in the smeared case is that the physical interpretation of $\mathbf{T}$ requires further clarification, and all formulas using $\mathbf{T}$ must be adapted (for example, when deriving the fragment-normalized energy).

In practice, several strategies are employed to improve efficiency and numerical stability. First, as $\mathbf{\Pi}(\omega)$ approaches zero for large $\omega$, its inversion becomes singular. To remedy this, Tikhonov regularization~\cite{hiltRidgeComputer1977} is applied so that the inversion is replaced by solving the eigenvalue equation
\begin{equation}\label{eq:regularizer}
\bigl[\mathbf{\Pi}^2(\omega) + \alpha \mathbf{1}\bigr]\,\mathbf{X}(\omega) = \mathbf{1}, \quad \alpha = e^{-12}
\end{equation}
which approximates $\mathbf{\Pi}^{-2}(\omega) \approx \mathbf{X}(\omega)$.

Second, the direct computation of the logarithm of a matrix is computationally expensive. Therefore, an eigenvalue decomposition of $\mathbf{1}-\mathbf{\Pi}(\omega)$ is performed prior to applying the logarithm:
\begin{equation}\label{eq:eigendecomp}
\mathbf{1} - \mathbf{\Pi}(\omega) = \mathbf{U}(\omega)\,\lambda(\omega)\,\mathbf{U}^{\dagger}(\omega)
\end{equation}
\begin{equation}\label{eq:log_decomp}
\ln\bigl(\mathbf{1} - \mathbf{\Pi}(\omega)\bigr) = \mathbf{U}(\omega)\,\ln\bigl(\lambda(\omega)\bigr)\,\mathbf{U}^{\dagger}(\omega)
\end{equation}
where $\mathbf{U}(\omega)$ is the unitary eigenvector matrix and $\lambda(\omega)$ are the eigenvalues.

\paragraph{Fragment energy functional}

We have shown that one can compute the double excitation amplitudes $\mathbf{T}$ exactly from an ACFDT-dRPA calculation in a non-iterative manner for a smeared mean-field (HF or DFT). This non-iterative method of obtaining cluster amplitudes from ACFDT-RPA motivates our current work. However, when computing the fragment energy from $\mathbf{T}$, the fragment energy functional used in all local/embedding methods (e.g., LNO-CC, SIE-CC, etc.) relies on the fact that the energy is invariant with respect to unitary rotations performed separately on the occupied and virtual spaces. In a smeared mean-field, the presence of fractionally occupied orbitals—mixtures of fully occupied and virtual states—complicates this invariance: simply rotating the fully and partially occupied orbitals together can alter the wavefunction and, consequently, the energy functional. Our goal is to derive an energy functional suitable for a smeared dRPA embedding.

For the trivial case where the embedding orbitals are either fully occupied or fully empty, the spin-restricted dRPA correlation energy is computed as
\begin{equation}
E_c^{\text{dRPA}} = t_{ij}^{ab}\,(ia|jb)
\end{equation}
Assuming the dRPA calculation is performed on an embedding Hamiltonian, the fragment correlation energy is given by
\begin{equation}
E_{c,F}^{\text{dRPA}} = t_{\tilde{i}\tilde{j}}^{ab}\,(\tilde{i}a|\tilde{j}b)
\end{equation}
with the transformed amplitudes defined as
\begin{equation}
t_{\tilde{i}\tilde{j}}^{ab} = \tfrac{1}{2}\Bigl[P^F_{\tilde{i}i}\, t_{ij}^{ab} + P^F_{\tilde{j}j}\, t_{ij}^{ab}\Bigr]
\end{equation}
where the projection operator $\hat{P}^F$ extracts the contribution of fragment $F$ orbitals (denoted by $x$) from the embedding occupied orbital $i$ using the coefficient matrix in the embedding basis:
\begin{equation}
P^F_{\tilde{i}i} = C_{\tilde{i}x}^{\dagger}\, C_{xi}
\end{equation}

This transformation is straightforward for the fully occupied space. However, when orbitals have fractional occupancies, the same formula cannot be applied. Recall that natural orbitals diagonalize the one-particle density matrix (1-RDM) with eigenvalues equal to the occupancies:
\begin{equation}
\begin{bmatrix} \mathbf{C}_{\text{o}} & \mathbf{C}_{\text{f}} & \mathbf{C}_{\text{v}} \end{bmatrix}^{\dagger}
\mathbf{D}\,
\begin{bmatrix} \mathbf{C}_{\text{o}} & \mathbf{C}_{\text{f}} & \mathbf{C}_{\text{v}} \end{bmatrix}
=
\begin{bmatrix} \mathbf{2} & \mathbf{f} & \mathbf{0} \end{bmatrix}
\end{equation}
Here, $\mathbf{C}_{\text{o}}$, $\mathbf{C}_{\text{f}}$, and $\mathbf{C}_{\text{v}}$ denote the fully occupied, fractionally occupied, and virtual orbitals, respectively; $\mathbf{D}$ is the 1-RDM; $\mathbf{2}$ represents eigenvalues of two (doubly occupied), $\mathbf{f}$ are fractional occupancies, and $\mathbf{0}$ represents eigenvalues of zero (empty).  

Unitary rotation of the fully occupied space $\mathbf{C}_{\text{o}}$ or the virtual space $\mathbf{C}_{\text{v}}$ preserves the eigenvalues:
\begin{equation}
\mathbf{U}_{\text{o}}^{\dagger}\,\mathbf{C}_{\text{o}}^{\dagger}\mathbf{D}\,\mathbf{C}_{\text{o}}\,\mathbf{U}_{\text{o}} = \mathbf{2}
\end{equation}
\begin{equation}
\mathbf{U}_{\text{v}}^{\dagger}\,\mathbf{C}_{\text{v}}^{\dagger}\mathbf{D}\,\mathbf{C}_{\text{v}}\,\mathbf{U}_{\text{v}} = \mathbf{0}
\end{equation}
However, unitary rotation of the fractionally occupied space $\mathbf{C}_{\text{f}}$ by $\mathbf{U}_{\text{f}}$ does not yield a diagonal representation of the 1-RDM $\mathbf{D}$ because it mixes eigenvectors with different eigenvalues:
\begin{equation}
\mathbf{U}_{\text{f}}^{\dagger}\,\mathbf{C}_{\text{f}}^{\dagger}\mathbf{D}\,\mathbf{C}_{\text{f}}\,\mathbf{U}_{\text{f}} \neq \mathbf{f}
\end{equation}

In the case of a smeared mean-field, indices $i$ or $j$ in $t_{ij}^{ab}$ may correspond to either fully occupied or partially occupied orbitals. We thus seek to derive a projection operator $\mathbf{P}^F$ that meets two criteria:
\begin{enumerate}
    \item It only mixes fully occupied orbitals among themselves.
    \item It rotates each partially occupied orbital into its fragment space without mixing it with other orbitals; i.e., off-diagonal contributions must be zero.
\end{enumerate}
A projection operator with these properties can be written in block-diagonal form:
\begin{equation}
\mathbf{P}^F = \begin{bmatrix} \mathbf{P}^F_{\text{o}} & 0 \\[5mm] 0 & \mathbf{P}^F_{\text{f}} \end{bmatrix}
\end{equation}
with
\begin{equation}
\left(\mathbf{P}^F_{\text{o}}\right)_{\tilde{i}i} = C_{\tilde{i}x}^{\dagger}\, C_{xi} \quad \text{(for fully occupied orbitals)},
\end{equation}
and
\begin{equation}
\left(\mathbf{P}^F_{\text{f}}\right)_{\tilde{i}i} =
\begin{cases}
C_{\tilde{i}x}^{\dagger}\, C_{xi}, & \text{if } \tilde{i} = i \\
0, & \text{if } \tilde{i} \neq i
\end{cases}
\end{equation}
for fractionally occupied orbitals.

It should be noted that this choice of projection is not unique. In general, the projections onto the fragment space must be complete, meaning that the sum over the projected electron density should equal the total electron density of the system—a property that requires both formal proof and numerical validation.

This discussion sets the stage for constructing an energy functional tailored for a smeared dRPA embedding, accommodating the complications introduced by fractional orbital occupancies.

\subsubsection{High-level AFQMC impurity solver}

Here we discuss how to obtain the energy for a fragment within an AFQMC run. It follows from the local energy expressions that within an embedding space one needs only restrict the summation over atomic orbital (AO) indices to the fragment (denoted by $x$) to compute fragment energies. This is formally analogous to the RDM-based energy formula in DMET, where only fragment indices are summed over in the energy equation, and contrasts with LNO-AFQMC, which requires a rotation into the excitation space. The choice of fragment energy functional is not unique.

Generally, the AFQMC local correlation energy for a walker $z$ can be written as
\begin{equation}
E_{\mathrm{corr}}^{z} = \sum_{i,j,a,b} G_{ia}^{z} G_{jb}^{z} \, [ 2(ia|jb) - (ib|ja) ]
\end{equation}
where $G_{ia}^{z}$ is the correlated one-body Green's function for walker $z$, and $(ia|jb)$ are the standard two-electron integrals in the excitation space.

To obtain a form valid for both gapped and gapless systems, irrespective of the mean-field gap, it can be rewritten as
\begin{equation}
E_{\mathrm{corr}}^{z} = \sum_{i,j,a,b} (G_{ia}^{z} - G^{0}_{ia})(G_{jb}^{z} - G^{0}_{jb}) \, [ 2(ia|jb) - (ib|ja) ]
\label{eq:gapped_ecorr}
\end{equation}
where $G^{0}_{ia}$ is the mean-field one-body Green's function, which is strictly zero for a gapped system.

The smeared Green's function is defined as
\begin{equation}
G_{pq}^{z} = \big( \mathbf{C}_{\phi_z} \, (\mathbf{C}_{\psi_o}^\dagger \mathbf{C}_{\phi_z})^{-1} \, \mathbf{f}_{\psi_o} \, \mathbf{C}_{\psi_o}^\dagger \big)_{qp}
\end{equation}
where $\mathbf{C}_{\phi_z}$ is the walker orbital coefficient matrix, $\mathbf{C}_{\psi_o}$ is the trial orbital coefficient matrix, and $\mathbf{f}_{\psi_o}$ is the diagonal occupation matrix of the smeared trial.

For gapless systems, the occupied indices $i$ include both fully occupied and fractionally occupied orbitals, while the virtual indices $a$ include both fractionally occupied and fully virtual orbitals. In this case, In this case, the quadratic form in Eq.~\eqref{eq:gapped_ecorr} does not fully subtract the mean-field contribution associated with the smeared trial density. An additional smeared mean-field subtraction term $E_{\mathrm{mf-sub}}^{z}$ is therefore required to ensure consistency with the gapped limit and to recover the correct mean-field cancellation, leading to:
\begin{equation}
E_{\mathrm{mf-sub}}^{z} = - \sum_{i,a} (G_{ia}^{z} - G^{0}_{ia}) \, F^{0}_{ia}
\end{equation}
where $F^{0}_{ia}$ is the mean-field Fock matrix element. This term vanishes for gapped systems since $G^{0}_{ia} = 0$, and and the off-diagonal occupied–virtual Fock elements vanish in the canonical basis $F^{0}_{ia} = 0$, but is non-zero for gapless systems and must be subtracted from Eq.~\eqref{eq:gapped_ecorr}.

The working expression for the local correlation energy with smearing is therefore
\begin{equation}
E_{\mathrm{corr}}^{z} = \sum_{i,j,a,b} (G_{ia}^{z} - G^{0}_{ia})(G_{jb}^{z} - G^{0}_{jb}) \, [ 2(ia|jb) - (ib|ja) ] 
+ \sum_{i,a} (G_{ia}^{z} - G^{0}_{ia}) \, F^{0}_{ia}
\label{eq:gapped_ecorr_mfsub}
\end{equation}

The fragment local correlation energy is obtained by replacing the full-system
Green’s function element $G_{ia}^z$ appearing in Eq.~\eqref{eq:gapped_ecorr_mfsub}
with its fragment-projected counterpart $G_{\tilde{i}a}^z$, defined via the
fragment projector as

\begin{equation}
G_{\tilde{i}a}^{z} = (\mathbf{P}^F_f)_{\tilde{i}i} \, G_{ia}^{z}
\end{equation}
where $\mathbf{P}^F_f$ is the fragment projection operator acting on fractionally occupied orbitals.

\clearpage

\section{Computational details}
\subsection{Cohesive energy}

The procedures employed for calculating and extrapolating the structural and properties of bulk BCC lithium (Li) and FCC aluminum (Al) closely follow the methodology outlined in previous work~\cite{Verena_LiAl}, to minimize errors arising from finite k-point sampling and basis set incompleteness. The following sections describe in detail the computational steps and parameters used for mean-field and correlated many-body electronic structure calculations, both for bulk systems and isolated atoms.

\subsubsection{Cohesive energy for BCC Li}
\paragraph*{Mean-field calculation for bulk Li}
For BCC Li, the isotropic primitive cell containing 2 Li atoms is used. The GTH-PADE pseudopotential is employed~\cite{GTHpseudo1,GTHpseudo2}.  
Mean-field calculations were performed with a double-zeta (DZ) basis using Monkhorst–Pack k-meshes ranging from $2 \times 2 \times 2$ to $8 \times 8 \times 8$. Extrapolation to the thermodynamic limit (TDL) was carried out using a linear inverse relation with respect to the total number of k-points:
\begin{equation}\label{eq:linear_inverse_relation}
E(N_k) = \frac{a}{N_k} + b,
\end{equation}
where $b$ is the extrapolated energy in the TDL. For HF, four $\mathbf{k}$-meshes—$5\times5\times5$, $6\times6\times6$, $7\times7\times7$, and $8\times8\times8$—were used to obtain $E_{\mathrm{HF}}(\mathrm{DZ/TDL})$.

Triple-zeta (TZ) and quadruple-zeta (QZ) calculations were performed with a fixed $6\times6\times6$ $\mathbf{k}$-mesh. The TDL energies for larger basis sets were evaluated via a compositional scheme:
$$
E_{\mathrm{HF}}(X\mathrm{Z/TDL}) = E_{\mathrm{HF}}(\mathrm{DZ/TDL}) + \Delta E_{\mathrm{HF}}(X\mathrm{Z}),
$$
where $\Delta E_{\mathrm{HF}}(X\mathrm{Z}) = E_{\mathrm{HF}}(X\mathrm{Z}, N_k=6^3) - E_{\mathrm{HF}}(\mathrm{DZ}, N_k=6^3)$. The QZ basis is taken to approximate the complete basis set (CBS) limit and define $E_{\mathrm{HF}}(\mathrm{CBS/TDL}) = E_{\mathrm{HF}}(\mathrm{QZ/TDL})$.

\paragraph*{Correlation calculation for bulk Li}
For correlated calculations using the random phase approximation (RPA), we applied the same TDL extrapolation scheme with a DZ basis using $\mathbf{k}$-meshes of $2\times2\times2$ and $3\times3\times3$. The extrapolated correlation energy $E_{\mathrm{c}}(\mathrm{DZ/TDL})$ was obtained using a two-point fit.

TZ and QZ basis calculations were performed with a fixed $3\times3\times3$ $\mathbf{k}$-mesh, and the corresponding TDL values were constructed as:
$$
E_{\mathrm{c}}(X\mathrm{Z/TDL}) = E_{\mathrm{c}}(\mathrm{DZ/TDL}) + \Delta E_{\mathrm{c}}(X\mathrm{Z}),
$$
with $\Delta E_{\mathrm{c}}(X\mathrm{Z}) = E_{\mathrm{c}}(X\mathrm{Z}, N_k=3^3) - E_{\mathrm{c}}(\mathrm{DZ}, N_k=3^3)$.

To reach the CBS limit, results from TZ and QZ basis sets were extrapolated using the following formula~\cite{CBSextrap}:
$$
E(X) = \frac{a}{X^3} + b, \quad X = 3, 4,
$$
where $b$ yields $E_{\mathrm{c}}(\mathrm{CBS/TDL})$.

An energy correction was also applied as a rigid shift~\cite{Verena_LiAl} using the difference:
$$
\Delta E'_{\mathrm{c}}(\mathrm{DZ}) = E_{\mathrm{c}}(\mathrm{DZ}, N_k=4^3) - E_{\mathrm{c}}(\mathrm{DZ}, N_k=3^3),
$$
evaluated at a lattice constant of 3.50 Å. The corrected energy is then
$$
E'_{\mathrm{c}}(\mathrm{CBS/TDL}) = E_{\mathrm{c}}(\mathrm{CBS/TDL}) + \Delta E'_{\mathrm{c}}(\mathrm{DZ}).
$$
    
\paragraph*{Single atom calculation}
To correct for basis set superposition error (BSSE) in cohesive energy calculations, ghost atoms were placed according to the bulk atomic geometry.  
For mean-field calculations on an isolated Li atom, six shells of ghost atoms (64 total) were included. The ghost atoms are placed at the same positions as in the crystal. Open boundary conditions (OBC) are used for the atomic calculation. Calculations were performed with DZ, TZ, and QZ basis sets, with QZ taken to approximate the CBS limit.  
Many-body electronic calculations of RPA and AFQMC included two shells of ghost atoms (14 total). Correlation energies were extrapolated to the CBS limit using TZ and QZ basis sets via $E(X) = aX^{-3} + b$.

\paragraph*{Embedding calculations}
The lattice constant, bulk modulus, and cohesive energy were obtained by fitting a Birch–Murnaghan equation~\cite{BMequation}. Embedding calculations were performed at the equilibrium lattice constant. The embedding AFQMC calculations were carried out at $\mathbf{k}$-meshes of $2\times2\times2$ and $3\times3\times3$ using a DZ basis, and $\mathbf{k}$-meshes of $3\times3\times3$ with TZ and QZ bases.  
The correlation energy was extrapolated using a linear model based on RPA energies:
$$
E_{\mathrm{QMC}} = k E_{\mathrm{RPA}} + b.
$$
The fitted parameters $k$ and $b$ were then applied to the full system energy:
$$
E_{\mathrm{QMC}}^{\mathrm{full}} = k E_{\mathrm{RPA}}^{\mathrm{full}} + b.
$$
The same TDL extrapolation scheme of bulk RPA with a DZ basis using $\mathbf{k}$-meshes of $2\times2\times2$ and $3\times3\times3$ is employed. The CBS limit is extrapolated with a $3\times3\times3$ $\mathbf{k}$-mesh using TZ and QZ bases. To further reduce the error from finite $\mathbf{k}$-meshes, the rigid shift correction from bulk Li RPA is applied~\cite{Verena_LiAl}.

\subsubsection{Cohesive energy for FCC Al}
The same computational protocol as for BCC Li is used for FCC Al. In the following, we briefly list the procedure with differences in parameter choices.

\paragraph*{Mean-field calculation for bulk Al}
For FCC Al, the isotropic cubic cell containing 4 Al atoms is used. The GTH-HF pseudopotential~\cite{GTHpseudo1,GTHpseudo2} is employed, following the benchmark in Ref.~\cite{Verena_LiAl}.  
To obtain $E_{\mathrm{HF}}(\mathrm{DZ/TDL})$, the same TDL extrapolation scheme as for BCC Li is used for $\mathbf{k}$-meshes from $5\times5\times5$ to $8\times8\times8$ with a DZ basis. Similarly, the TDL energies for larger basis sets are evaluated via the compositional scheme using a $5\times5\times5$ $\mathbf{k}$-mesh for TZ and a $4\times4\times4$ $\mathbf{k}$-mesh for QZ.

\paragraph*{Correlation calculation for bulk Al}
For correlation calculations, the anisotropic primitive cell containing 2 Al atoms is used. $E_{\mathrm{c}}(\mathrm{DZ/TDL})$ is estimated with a $5\times5\times2$ $\mathbf{k}$-mesh, followed by a rigid shift~\cite{Verena_LiAl}.  
For correlated calculations using RPA, we applied the exact linear inverse TDL extrapolation scheme as used for HF with a DZ basis using $\mathbf{k}$-meshes of $2\times2\times2$ and $3\times3\times3$. The extrapolated correlation energy $E_{\mathrm{c}}(\mathrm{DZ/TDL})$ was obtained using a two-point fit.

TZ and QZ basis calculations were performed with a fixed $3\times3\times3$ $\mathbf{k}$-mesh, and the corresponding TDL values were constructed as:
$$
E_{\mathrm{c}}(X\mathrm{Z/TDL}) = E_{\mathrm{c}}(\mathrm{DZ/TDL}) + \Delta E_{\mathrm{c}}(X\mathrm{Z}),
$$
with $\Delta E_{\mathrm{c}}(X\mathrm{Z}) = E_{\mathrm{c}}(X\mathrm{Z}, N_k=3^3) - E_{\mathrm{c}}(\mathrm{DZ}, N_k=3^3)$.

To reach the CBS limit, results from TZ and QZ basis sets were extrapolated using the following formula~\cite{CBSextrap}:
$$
E(X) = \frac{a}{X^3} + b, \quad X = 3, 4,
$$
where $b$ yields $E_{\mathrm{c}}(\mathrm{CBS/TDL})$.

An energy correction was applied as a rigid shift~\cite{Verena_LiAl} using the difference:
$$
\Delta E'_{\mathrm{c}}(\mathrm{DZ}) = E_{\mathrm{c}}(\mathrm{DZ}, N_k=4^3) - E_{\mathrm{c}}(\mathrm{DZ}, N_k=3^3),
$$
evaluated at a lattice constant of 3.50 Å. The corrected energy is then
$$
E'_{\mathrm{c}}(\mathrm{CBS/TDL}) = E_{\mathrm{c}}(\mathrm{CBS/TDL}) + \Delta E'_{\mathrm{c}}(\mathrm{DZ}).
$$

\paragraph*{Single atom calculation}
The same procedure as for Li single-atom calculations is applied to Al.  
For mean-field calculations on an isolated Al atom, six shells of ghost atoms (86 total) were included.  
Many-body electronic calculations of RPA and AFQMC included two shells of ghost atoms (18 total).

\subsection{CO adsorption on the Cu(111) surface}

\subsubsection{Geometry optimizations}
All geometry optimizations were performed using the Perdew–Burke–Ernzerhof~(PBE) exchange-correlation functional~\cite{PBE} within the plane-wave density functional theory~(DFT) framework, as implemented in the Vienna Ab Initio Simulation Package~(VASP)~\cite{VASP}. The Kohn–Sham wavefunctions were expanded in a plane-wave basis set with a kinetic energy cutoff of 600 eV, and core–electron interactions were treated using the projector augmented-wave (PAW) method~\cite{PAW}.

For bulk copper, a $\Gamma$-centered Monkhorst–Pack grid~\cite{monkhorstSpecialPoints1976} of $12 \times 12 \times 12$ was used for Brillouin zone sampling. The optimized lattice constant of 3.632~\AA\ is in good agreement with the experimental value of 3.615~\AA~\cite{arblasterSelectedValues2018}.

Surface systems were modeled using a four-layer $2 \times 2$ Cu(111) slab with periodic boundary conditions, separated by a 10~\AA\ vacuum layer to minimize interactions between periodic images. Structural relaxations were performed using a $\Gamma$-centered $7 \times 7 \times 1$ k-point mesh, with the bottom two layers fixed during optimization. Convergence criteria were set to 0.01 eV/\AA\ for residual forces and $10^{-6}$ eV for total energy self-consistency.

\subsubsection{Computational settings for electronic structure calculations}
The mean-field calculations for surface adsorption systems were carried out using the PySCF~\cite{PySCF_1,PySCF_2} package, based on geometries optimized with VASP~\cite{VASP}. The def2-SVP and def2-DZVP basis sets were employed for all atoms~\cite{def2_1,def2_2}. For periodic boundary condition (PBC) calculations, diffuse functions with exponents smaller than 0.05 were excluded to improve numerical stability. Two-electron integrals were evaluated using range-separated Gaussian density fitting~\cite{RSDF}. Brillouin zone sampling was performed using a $\Gamma$-centered Monkhorst–Pack grid with $k$-meshes ranging from $3 \times 3 \times 1$ to $6 \times 6 \times 1$.

Hartree–Fock~(HF) and density functional approximations, including PBE~\cite{PBE} and PBE0~\cite{PBE0}, were used for reference mean-field calculations and in periodic RPA calculations. Embedding calculations were carried out on $3 \times 3 \times 1$ and $4 \times 4 \times 1$ $k$-meshes with the def2-SVP basis set, using the PBE functional~\cite{PBE} as the reference.

\subsubsection{Adsorption energy evaluation}\label{ads_ener}
The adsorption of a small molecule, denoted as $M$, onto a metallic surface modeled as a multilayer slab is defined as
\begin{equation}\label{eq:adsorption_energy1}
E_{\text{ads}} = E_{\text{M}+\text{slab}} - E_{\text{M}} - E_{\text{slab}}
\end{equation}
where $E_{\text{ads}}$ is the adsorption energy. $E_{\text{M}+\text{slab}}$ is the total energy of the combined system calculated in the geometry of the adsorbed state, while $E_{\text{slab}}$ and $E_{\text{M}}$ are the energies of the isolated slab and the isolated molecule, respectively, calculated in their own equilibrium geometries.

In practice, it is numerically advantageous to compute the interaction energy ($E_{\text{int}}$), which has the same form as the adsorption energy but with each term calculated in the geometry of the adsorbed state. The differences arising from the geometries of the isolated species and their adsorbed configurations are corrected separately. This additional term, denoted $E_{\text{reorg}}$, is defined as the energy difference arising from geometry relaxation to the adsorption energy:
\begin{equation}\label{eq:adsorption_energy2}
E_{\text{ads}} = E_{\text{int}} + E_{\text{reorg}}
\end{equation}

The use of the same geometry for all three components in the interaction energy equation permits the application of the counterpoise procedure—a common BSSE correction method—to mitigate basis set superposition error in GTO calculations~\cite{Boys01101970}. In particular, the BSSE-corrected interaction energy is computed as
\begin{equation}\label{eq:interaction_energy_cp}
E_{\text{int}} = E_{\text{M}+\text{slab}}^{\text{M}+\text{slab}} - E_{\text{M}}^{\text{M}+\text{slab}} - E_{\text{slab}}^{\text{M}+\text{slab}}
\end{equation}
where $E_{\text{M+slab}}^{\text{M+slab}}$ is the energy of the composite system computed in the composite basis set, $E_{\text{M}}^{\text{M+slab}}$ is the energy of the isolated molecule computed with ghost basis functions for the slab, and $E_{\text{slab}}^{\text{M+slab}}$ is the energy of the isolated slab computed with ghost basis functions for the molecule.

For CO on the copper (111) surface, the interaction energy is defined as
\begin{equation}\label{eq:interaction_energy_CO_Cu}
E_{\text{int}} = E_{\text{CO}+\text{Cu}}^{\text{CO}+\text{Cu}} - E_{\text{CO}}^{\text{CO}+\text{Cu}} - E_{\text{Cu}}^{\text{CO}+\text{Cu}}
\end{equation}
The details of how the slab model for Cu(111) is defined and how $E_{\text{reorg}}$ is computed will be discussed in Section~\ref{method_assessment}.

For correlated calculations, the total energy consists of two contributions: the mean-field energy and the correlation energy. Similarly, the interaction energy can be decomposed into these components. This approach not only aids in understanding the nature of adsorption but also provides a framework for incorporating correction terms within composite methods. In particular, the following decomposition is employed:
\begin{equation}\label{eq:interaction_energy_decomposed}
E_{\text{int}} = E_{\text{int}}^{\text{MF}} + E_{\text{int}}^{\text{corr}} 
\end{equation}
where $E_{\text{int}}^{\text{MF}}$ is the interaction energy computed at the mean-field level (e.g., Hartree–Fock or DFT), and $E_{\text{int}}^{\text{corr}}$ is the correlation energy obtained using a correlated method such as dRPA, CCSD, or AFQMC.

\subsection{\ce{H2} desorption energy barrier}
\subsubsection{Computational settings for electronic structure calculations}

The electronic structure calculations for the reactant and transition states were performed in PySCF~\cite{PySCF_1,PySCF_2}, based on the geometric structures from Ref.~\citenum{zhouLightdrivenMethane2020}. The def2-SVP basis set~\cite{def2_1,def2_2} was employed for all atoms, with diffuse functions having exponents smaller than 0.05 excluded. Two-electron integrals were evaluated using range-separated Gaussian density fitting~\cite{RSDF}. Brillouin zone sampling was performed using a $\Gamma$-centered Monkhorst–Pack grid with $k$-meshes ranging from $2 \times 2 \times 1$ to $5 \times 5 \times 1$.
Embedding calculations were carried out on $2 \times 2 \times 1$ and $4 \times 4 \times 1$ $k$-meshes with the def2-SVP basis set, using the orbitals calculated from PBE functional~\cite{PBE} as the reference. For all embedding calculations, the external bath was constructed using approximate MP2 with orbital thresholds of  $\varepsilon_{occ}=0.05$ for occupied orbitals and $\varepsilon_{vir}=0.005$ for virtual states. We systematically included more atoms as the fragment corrected by the high-level solver. The fragment was defined in successive steps: initially as the two H atoms~(\ce{H2}), then expanded to include the two nearest Cu atoms~\ce{H2Cu2}, and finally expanded further to include the four nearest Cu atoms~\ce{H2Cu4}. The detailed results of the embedding is list in Table~\ref{tab:emb_h2}.
\clearpage

\subsection{Adsorption on 3d transition metal doped single atom alloy}
\subsubsection{Computational settings for electronic structure calculations}
\label{subsec:SAA}
The electronic structure calculations were performed using the PySCF~\cite{PySCF_1,PySCF_2}. The single-atom alloy (SAA) systems were modeled as three-layer slabs, with atomic coordinates adopted directly from the recent work of Schumann et al~\cite{schumannTenelectronCountRule2024}. Our baseline DFT calculations successfully reproduced the qualitative adsorption trends reported in the reference study. 
The nonlocal optB86b-vdW functional~\cite{klimesChemicalAccuracyVan2009,klimesVanWaalsDensity2011}, consistent with the reference work~\cite{schumannTenelectronCountRule2024}, is employed for all meanfield and post-meanfield calculations. For the 3d-doped SAA systems, spin-restricted calculations were employed in accordance with the reference study, yielding adsorption trends consistent with those reported in Fig. S4 from Ref.~\citenum{schumannTenelectronCountRule2024}. 
The def2-SVP basis set was employed for all atoms, with diffuse functions having exponents smaller than 0.05 excluded. Two-electron integrals were evaluated using range-separated Gaussian density fitting~\cite{RSDF}. For SAA systems, the Brillouin zone was sampled using a $\Gamma$-centered $2 \times 2 \times 1$ Monkhorst–Pack grid. As the primary focus of this study is the relative adsorption trend across the 3d transition-metal series, rather than absolute adsorption energies, this k-mesh was found to be sufficient for capturing the qualitative behavior of interest.  We validated this computational setup by confirming that it successfully reproduces the trends reported in the original work. 
The energies of isolated atoms (O, N, C) were derived from a combination of calculated molecular energies and experimental atomization energies obtained from the CCCBDB database~\cite{CCCBDB}. Specifically, the energy of an isolated oxygen or nitrogen atom was calculated as:
$E_{\text{O/N}} = \frac{1}{2}(E_{\text{O}_2/\text{N}_2} - E_{\text{a}})$, where $E_a$ represents the atomization energy. Similarly, the energy of an isolated carbon atom was determined using the calculated energy of CO and the previously derived energy of the oxygen atom:
$E_{\text{C}} = E_{\text{CO}} - E_{\text{O}} - E_{\text{a}}$.

For all embedding calculations, the number of external bath constructed using approximate MP2 was fixed as 150 orbitals, with half of them occupied, for both the bare slab and the adsorbate-slab systems. 
To focus the high-level computational effort on the center of chemical activity and the trend across the dopant series, the embedding fragment was defined as the adsorbate and the 3d transition-metal dopant atom. The surrounding host Cu atoms were therefore excluded from the high-level fragment calculation and were described by the global RPA treatment.

\clearpage

\section{Additional data}\label{method_assessment}
\subsection{Testing the virtual-to-occupied cutoff ratios}

To assess the robustness of the embedding RPA procedure with respect to virtual-space truncation, we compare embedding RPA and conventional RPA correlation energies for 1D Li ($6\times1\times1$) as a function of the occupied-orbital threshold $\varepsilon_{\mathrm{occ}}$, while varying the virtual-to-occupied cutoff ratio $\varepsilon_{\mathrm{vir}}/\varepsilon_{\mathrm{occ}}$.

\begin{figure*}[ht]
    \centering
    \includegraphics[width=0.85\linewidth]{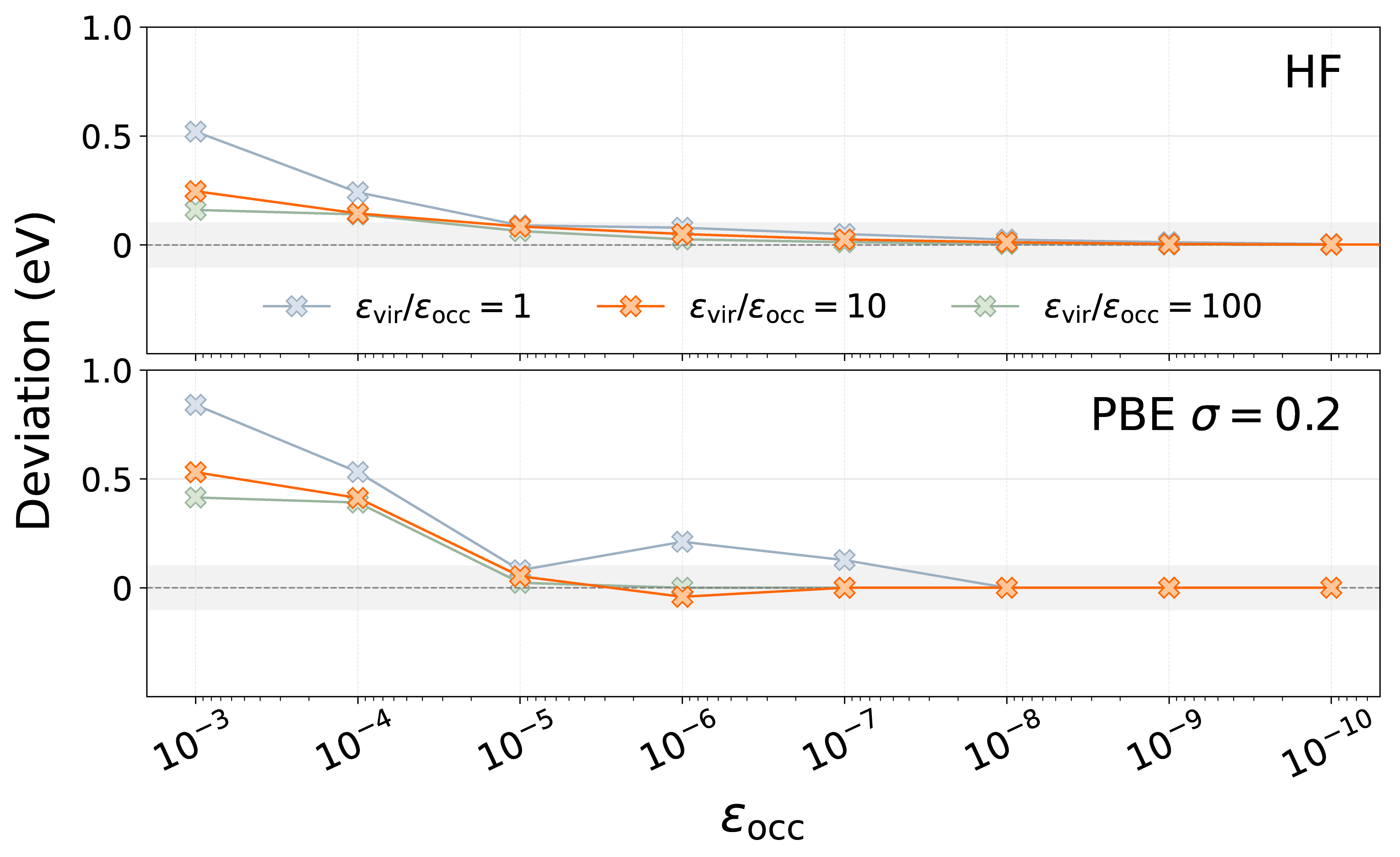}
    \caption{Deviations of embedding RPA for 1D lithium ($6\times1\times1$) from its conventional counterpart for HF and PBE ($\sigma=0.2$) mean fields as a function of $\varepsilon_{\mathrm{occ}}$, evaluated using virtual-to-occupied cutoff ratios $\varepsilon_{\mathrm{vir}}/\varepsilon_{\mathrm{occ}}$ of 1, 10, and 100. The deviations decrease systematically as the cutoff is tightened, indicating convergence toward the conventional RPA limit within the tested range.}
  \label{fig:threshold_testing}
\end{figure*}
\clearpage

\subsection{Testing the effect of thermal smearing on AFQMC convergence}

\begin{figure*}[ht]
    \centering
    \includegraphics[width=0.70\linewidth]{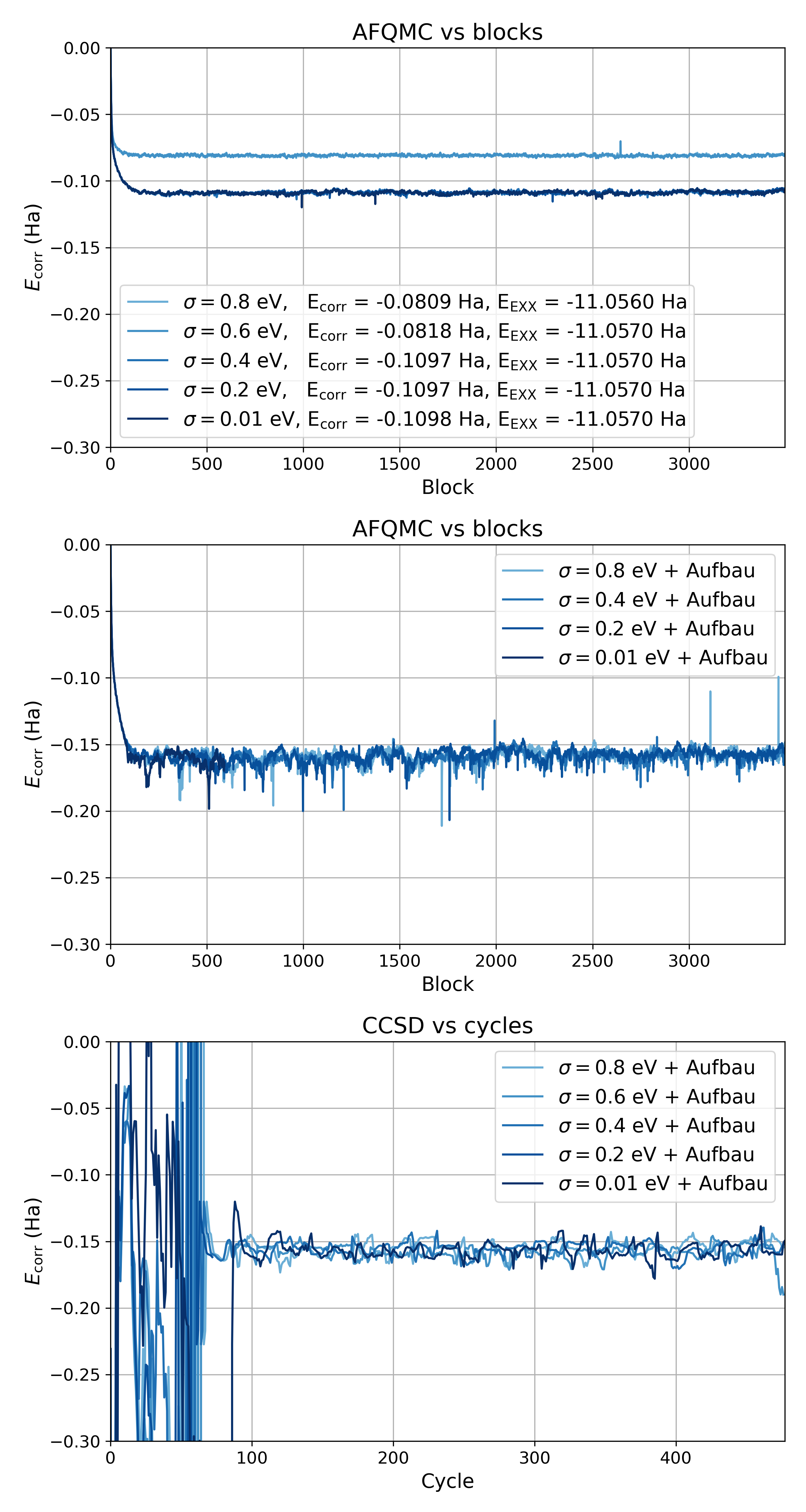}
    \caption{Convergence of AFQMC and CCSD correlation energies for bulk Li using thermally smeared mean-field trial references, comparing fractional-occupation and Aufbau-enforced trials.}
  \label{fig:AFQMC_vs_CCSD}
\end{figure*}

Fig.~\ref{fig:AFQMC_vs_CCSD} highlights the role of thermal smearing and trial-state construction in metallic many-body calculations. When fractional occupations from thermally smeared mean-field solutions are retained, AFQMC exhibits stable equilibration and reduced statistical fluctuations, with the correlation energy converging smoothly and showing only weak dependence on the smearing parameter over a broad range of $\sigma$ values. In contrast, enforcing Aufbau occupations on the same smeared mean-field orbitals reintroduces an effectively gapless trial reference, leading to pronounced fluctuations and numerical instabilities in the AFQMC energy estimator. A similar behavior is observed for CCSD calculations based on Aufbau references, which show severe oscillations and slow convergence. These results indicate that the numerical stability of correlated methods in metallic systems is strongly controlled by the presence of fractional occupations, rather than the specific choice of many-body solver.

\clearpage

\subsection{Li/Al results from literature and current work}

To benchmark the accuracy of the FEMION framework in a well-understood metallic regime, we compare cohesive energies, lattice constants, and bulk moduli for bulk Li and Al against established wavefunction-based and experimental reference data. These simple metals provide a stringent but controlled test case, as high-quality CCSD(T) and experimental benchmarks are available.

\begin{table}[h]
    \centering
    \begin{tabular}{lcccc}
        \hline
        \textbf{Li} & \textbf{a($\bf{\AA}$)}& \textbf{B(GPa)} & \textbf{$\bf{E\_{coh}}$(eV/atom)} & \textbf{Ref.} \\ \hline
        HF  & 3.68 & 9.0 & -0.599 & \citenum{Verena_LiAl}\\ 
        CCSD  & 3.50 & 12.5 & -1.388 & \citenum{Verena_LiAl}\\ 
        $\rm{CCSD(T)_{SR}}$ & 3.50 & 12.9 & -1.469 & \citenum{Verena_LiAl}\\ 
        RPA@PBE  & 3.47 & 13.2 & -1.388 & This work\\ 
        FEMION & --- & --- & -1.469 & This work\\ 
        Experiment & 3.45 & 13.3 & -1.66 &\citenum{kittelIntroductionSolid2018} \\ \hline
    \end{tabular}
    \bigskip
    \begin{tabular}{lcccc}
        \hline
        \textbf{Al} & \textbf{a($\bf{\AA}$)} & \textbf{B(GPa)} & \textbf{$\bf{E\_{coh}}$(eV/atom)}&\textbf{Ref.} \\ \hline
        HF  & 4.08 & 80.0 & -1.388 & \citenum{Verena_LiAl}\\ 
        CCSD  & 4.02 & 93.2 & -2.966 & \citenum{Verena_LiAl}\\ 
        CCSD(T)SR & 4.02 & 91.7 & -3.102 & \citenum{Verena_LiAl}\\ 
        RPA@PBE  & 4.02 & 92.1 & -3.075 & This work \\ 
        FEMION  & --- & --- & -3.075 & This work\\ 
        Experiment & 4.02 & 80.3 & -3.429 & \citenum{kittelIntroductionSolid2018} \\ \hline
    \end{tabular}
    
    \caption{Lattice parameter (\textit{a}), bulk modulus (\textit{B}), and cohesive energy (E\_coh) for Li and Al. The experimental value is corrected for zero-point motion (ZPM) using a ZPM correction obtained from Ref.~\citenum{zhangPerformanceVarious2018} via the HSE06 functional~\cite{HSE06}}
    \label{tab:li_al_properties}
\end{table}

Table~\ref{tab:li_al_properties} summarizes cohesive energies, lattice parameters, and bulk moduli for bulk Li and Al obtained using FEMION, alongside representative literature and experimental reference values. These data provide supporting benchmarks for the metallic cohesion results discussed in the main text.

\subsection{Computational efficiency for FEMION tested on Li bulk}

\begin{figure*}[ht]
    \centering
    \includegraphics[width=0.85\linewidth]{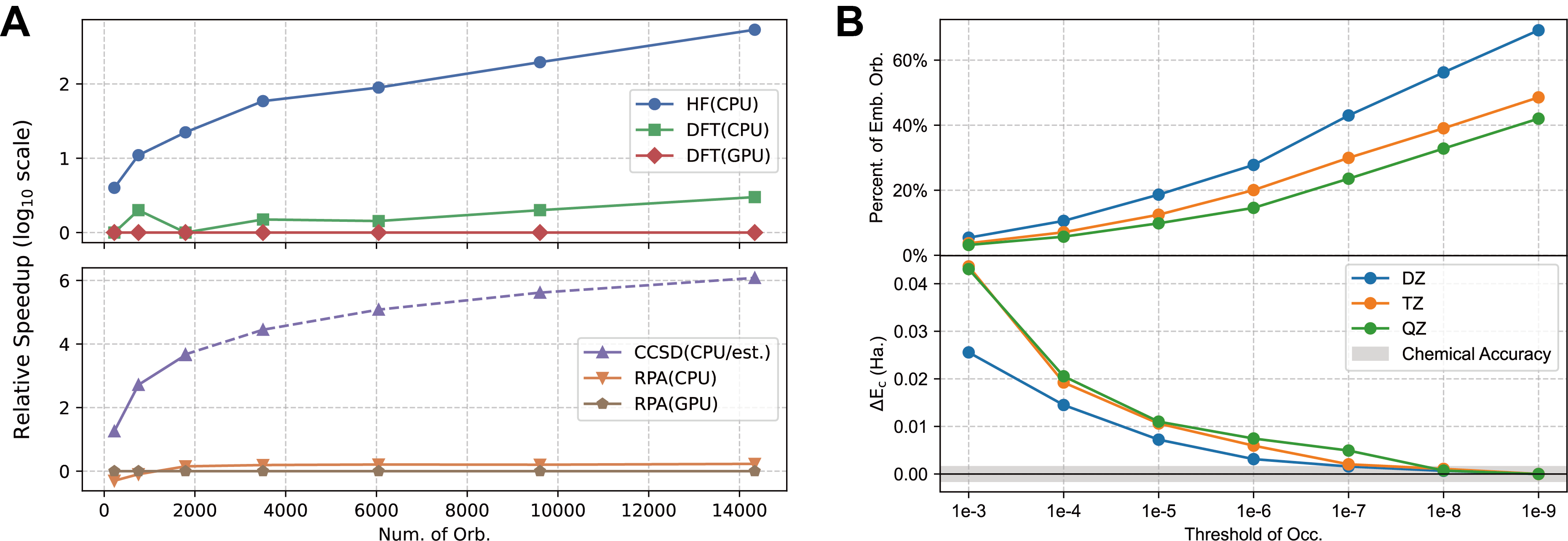}
    \caption{Computational efficiency and convergence of the FEMION framework.(a) Logarithmic relative speedup, defined as$\log_{10}(t/t_{ref})$, for calculations on bulk Li with varying k-point mesh densities. For mean-field and post-mean-field methods, the fastest GPU-accelerated calculations (DFT and RPA, respectively) serve as the reference ($t_{ref}$). The dashed line indicates estimated time for CCSD due to its prohibitive computational cost. (b) Top: Average size of the local natural orbital (LNO) fragment, expressed as a percentage of the total number of system orbitals. Bottom: Convergence of the AFQMC correlation energy with respect to the LNO threshold, referenced to the largest fragment calculation (threshold of occupied orbitals= $10^{-9}$)}
  \label{fig:Li_efficiency}
\end{figure*}

  Accurately treating metallic systems with correlated wavefunction methods has long been hindered by the absence of a band gap. The conventional strategy is to impose an artificial gap at the Hartree-Fock (HF) level and then apply post-HF methods such as CCSD. While systematically improvable, this approach is prohibitively expensive: HF itself is ill-conditioned under periodic boundary conditions due to costly exchange integrals, and CCSD scales as $O(N^6)$.  This cost is particularly crippling given that metallic systems require large supercells or dense k-point meshes to converge to the thermodynamic limit, rendering this entire approach intractable for most problems of chemical interest.
  FEMION circumvents these barriers by reformulating the framework to accommodate fractional occupations, enabling calculations to begin directly from thermally smeared DFT orbitals. This allows a smooth transition to periodic RPA calculations. The DFT+RPA routine yields cohesive energies that are consistent with established high-level reference trends for metallic systems such as Li and Al, while being substantially less expensive than HF-based coupled-cluster workflows on dense $k$-point meshes. To further improve accuracy beyond RPA, FEMION incorporates local correlation corrections from an AFQMC impurity solver, approaching the accuracy of high-level wavefunction methods for the systems considered here. While a full AFQMC calculation on the entire system remains intractable, the embedding formulation renders the correlated correction computationally tractable by restricting it to a localized fragment. The local correlation energy of the fragment converges rapidly, typically requiring less than 40\% of the total system's orbitals. Given the $O(N^3\text{--}N^4)$ scaling of AFQMC, restricting the correlated treatment to a compact fragment substantially reduces the effective computational cost relative to a full-system calculation, while retaining the dominant local correlation contributions.
  Together, these results illustrate how FEMION combines a scalable periodic RPA backbone with controllable local many-body corrections, enabling practical treatments of metallic cohesion and related surface chemistry problems.
  \clearpage
  \clearpage

\subsection{Geometry reconstruction of CO molecule}

As discussed in Section~\ref{ads_ener}, the CO bond length changes upon adsorption on the copper surface due to bond formation. A similar geometric reconstruction occurs for the copper slab, although to a much lesser extent, as previous work has shown its effect to be negligible.~\cite{NEEF20061085}

\begin{equation}\label{eq:reorg_energy}
E_{\text{reorg}} \approx E_{\text{CO}}^{\text{top/fcc}} - E_{\text{CO}}^{\text{PBE}}
\end{equation}

In this section, we investigate the impact of this reconstruction, which serves as a correction on atop of the computed interaction energy to yield the final adsorption energy. Specifically, we compute the energy difference between the CO molecule in the composite system (e.g., atop or fcc configuration) and its geometry in the isolated state, as obtained from PBE-level geometry optimizations reported in the literature. The energies are computed using both PBE and PBE0 functionals; the rationale behind these choices becomes clearer in the next section, where we discuss the construction of reliable slab models. Note that the energy of CO converges very quickly even with a 3x3x1 k-point meshes, although we employed an 11x11x1 meshes in our calculations. The data for the 11x11x1 meshes are presented in Table~\ref{tab:CO_DFT}.

\begin{table}[ht]
\centering
\caption{Geometry reorganization contribution (in eV) from CO using different DFT methods.}
\begin{tabular}{lccccc}
\hline
               & atop (1.158\AA) & fcc (1.182\AA) & PBE (1.135\AA) & $E_{\text{int}}$ atop & $E_{\text{int}}$ fcc \\
\hline
PBE            & -14.779443     & -14.711936    & -14.786532    & 0.007 &  0.075         \\
PBE0           & -19.205356     & -19.101772    & -19.245446    & 0.040 &  0.144         \\
\hline
\end{tabular}
\label{tab:CO_DFT}
\end{table}

\subsection{Slab models for CO@Cu(111)}

In this section, we discuss how many layers are sufficient to model the physics of CO on the Cu surface while achieving the chemical accuracy targeted in our work. We compare the adsorption energies obtained from a four-layer (4L) model and a five-layer (5L) model with literature results. The absolute interaction and adsorption energies can be found in the supplementary raw data; here, we present only the differences between our computed values and those reported in the literature to evaluate the models used. 

There are several observations from Table~\ref{tab:PBE_slab_models}. First, the deviation between the 4L and 5L models is roughly 0.043 eV (or 1 kcal/mol), although it can be slightly larger or smaller. In the case of the PBE functional, the deviation between the atop and fcc configurations is even smaller. For the PBE0 functional, the difference is slightly larger; however, the atop--fcc differences remain around 0.043 eV (or 1 kcal/mol). The larger deviation observed with PBE0 may be attributed to the exchange term, which is generally more challenging to handle in periodic boundary condition software. Overall, the 4L model strikes a good balance between accuracy and computational cost and will be used for all many-body calculations.

\begin{table}[ht]
\centering
\caption{Comparison of adsorption energies at PBE level for different k-point meshes and slab models. (MSE: mean signed error, MUE: mean unsigned error.)}
\label{tab:PBE_slab_models}
\begin{tabular}{l ccc ccc}
\hline
       & \multicolumn{3}{c}{5L} & \multicolumn{3}{c}{4L} \\
\cline{2-7}
$N_k \times N_k \times 1$      & $E_{atop}$ & $E_{fcc}$ & $E_{atop}-E_{fcc}$ & $E_{atop}$ & $E_{fcc}$ & $E_{atop}-E_{fcc}$ \\
\hline
$3 \times 3 \times 1$  & -0.10 & -0.15 & 0.05  & -0.22 & -0.27 & 0.05 \\
$5 \times 5 \times 1$  &  0.06 &  0.02 & 0.04  & -0.07 & -0.06 & -0.01 \\
$7 \times 7 \times 1$  & -0.04 & -0.03 & -0.01 & -0.07 & -0.02 & -0.05 \\
$9 \times 9 \times 1$  & -0.05 & -0.08 & 0.02  & -0.09 & -0.09 & 0.00 \\
$11 \times 11 \times 1$ &  0.01 &  0.02 & -0.01 & -0.08 & -0.05 & -0.04 \\
\hline
MSE    & -0.02 & -0.04 & 0.02  & -0.11 & -0.10 & -0.01 \\
MUE    &  0.05 &  0.06 & 0.03  &  0.11 &  0.10 & 0.03 \\
\hline
\end{tabular}
\end{table}

\begin{table}[ht]
\centering
\caption{Comparison of adsorption energies at PBE0 level for different k-point meshes and slab models. (MSE: mean signed error, MUE: mean unsigned error.)}
\label{tab:PBE0_slab_models}
\begin{tabular}{l ccc ccc}
\hline
       & \multicolumn{3}{c}{5L} & \multicolumn{3}{c}{4L} \\
\cline{2-7}
$N_k \times N_k \times 1$      & $E_{atop}$ & $E_{fcc}$ & $E_{atop}-E_{fcc}$ & $E_{atop}$ & $E_{fcc}$ & $E_{atop}-E_{fcc}$ \\
\hline
$3 \times 3 \times 1$  & -0.23 & -0.22 & -0.01    & -0.25 & -0.18 & -0.07    \\
$5 \times 5 \times 1$  &  0.03 &  0.06 & -0.03    & -0.05 &  0.05 & -0.10    \\
$7 \times 7 \times 1$  & -0.06 &  0.01 & -0.07    & -0.11 &  0.06 & -0.17    \\
$9 \times 9 \times 1$  & -0.07 & -0.03 & -0.04    & -0.09 &  0.00 & -0.09    \\
$11 \times 11 \times 1$  &  0.00 &  0.08 & -0.08    & -0.09 &  0.05 & -0.14    \\
\hline
MSE    & -0.07 & -0.02 & -0.05    & -0.12 &  0.00 & -0.11    \\
MUE    &  0.08 &  0.08 &  0.05    &  0.12 &  0.07 &  0.11    \\
\hline
\end{tabular}
\end{table}
\clearpage

\subsection{DFT cross-validation of GTO and planewave methods for CO@Cu(111)}

As periodic quantum chemistry for realistic solids is still in its infancy, the quality of widely used molecular Gaussian-type orbital (GTO) basis functions remains questionable. In many cases, these basis sets face severe numerical instabilities due to basis set superposition errors from diffuse functions. At the same time, pseudopotentials and solid-state optimized basis sets are primarily designed for DFT or HF computations and are not adequate for post-HF periodic quantum chemistry. While recent efforts~\cite{Hongzhou_GTO,Verena_LiAl} have been made to develop re-optimized molecular basis sets for correlated periodic calculations, their applications are still restricted to a limited number of elements, and the method remains under active development.

While the design of the next generation of correlated basis sets is critical for the future of periodic quantum chemistry, we have chosen a different approach at this time. We reuse molecular basis sets and remove only a minimal amount of diffuse functions. The resulting basis sets are then tested against plane-wave level calculations. Later, we also demonstrate their reliability at a more economical, DFT-correlated level—specifically, using RPA. This strategy ensures that our chosen basis functions are reliable for correlated computations on this system.

\begin{table}[h!t]
\centering
\caption{Comparison between computed interaction energies (in eV) for the atop configuration obtained using different methods. Results from PySCF with a Gaussian-type orbital (GTO) basis are compared with those from VASP using a plane-wave basis set at various k-point grids. Missing values are denoted by “--”.}
\label{tab:int_energy_pyscf_vasp_pbe}
\begin{tabular}{llccc}
\hline
\textbf{Grid} & \textbf{Method}         & \textbf{HF} & \textbf{PBE}  & \textbf{PBE0} \\
\hline
\multirow{5}{*}{3$\times$3$\times$1} 
 & def2-SVP         & 0.462         & -0.873         & -0.708 \\
 & def2-TZVP        & 0.492         & -0.914         & -0.769 \\
 \cline{2-5}
 & CBS              & 0.504         & -0.932         & -0.796 \\
 & VASP          & --            & -0.953         & -0.825 \\
 & Deviation  & --            & 0.021          & 0.029  \\
\hline
\multirow{5}{*}{4$\times$4$\times$1} 
 & def2-SVP         & 0.724         & -0.689         & -0.494 \\
 & def2-TZVP        & 0.747         & -0.744         & -0.560 \\
 \cline{2-5}
 & CBS              & 0.756         & -0.767         & -0.589 \\
 & VASP             & --            & -0.801         & -0.633 \\
 & Deviation        & --            & 0.034          & 0.044  \\
\hline
\multirow{4}{*}{5$\times$5$\times$1} 
 & def2-SVP         & --            & -0.670         & --     \\
 & def2-TZVP        & --            & -0.723         & --     \\
 \cline{2-5} 
 & CBS              & --            & -0.747         & --      \\
 & VASP             & --            & -0.789         & --      \\
 & Deviation        & --            & 0.042          & --     \\
\hline
\end{tabular}
\end{table}

\begin{table}[h!t]
\centering
\caption{Comparison between computed interaction energies (in eV) for the fcc configuration obtained using different methods. Results from PySCF with a Gaussian-type orbital (GTO) basis are compared with those from VASP using a plane-wave basis set at various k-point grids. Missing values are denoted by “--”.}
\label{tab:int_energy_pyscf_vasp_pbe0}
\begin{tabular}{llccc}
\hline
\textbf{Grid} & \textbf{Method}         & \textbf{HF} & \textbf{PBE}  & \textbf{PBE0} \\
\hline
\multirow{5}{*}{3$\times$3$\times$1} 
 & def2-SVP         & 1.336         & -1.102         & -0.750 \\
 & def2-TZVP        & 1.528         & -1.155         & -0.816 \\
 \cline{2-5}
 & CBS              & 1.612         & -1.178         & -0.844 \\
 & VASP             & --            & -1.234         & -0.905 \\
 & Deviation   & --            & 0.056          & 0.060  \\
\hline
\multirow{5}{*}{4$\times$4$\times$1} 
 & def2-SVP         & 1.473         & -0.918         & -0.589 \\
 & def2-TZVP        & 1.579         & -0.983         & -0.640 \\
 \cline{2-5}
 & CBS              & 1.625         & -1.011         & -0.662 \\
 & VASP             & --            & -1.083         & -0.774 \\
 & Deviation        & --            & 0.072          & 0.112  \\
\hline
\multirow{5}{*}{5$\times$5$\times$1} 
 & def2-SVP         & --            & -0.846         & --      \\
 & def2-TZVP        & --            & -0.899         & --     \\
 \cline{2-5}
 & CBS              & --            & -0.922         & --     \\
 & VASP             & --            & -1.004         & --     \\
 & Deviation        & --            & 0.082          & --     \\
\hline
\label{tab:meanfield_kmesh}
\end{tabular}
\end{table}

Table~\ref{tab:int_energy_pyscf_vasp_pbe} and Table~\ref{tab:int_energy_pyscf_vasp_pbe0} show that we achieve chemical accuracy agreement with VASP for the atop configuration for both the PBE and PBE0 functionals. For the fcc configurations, our results are underestimated compared to the VASP plane-wave results. These deviations will be interpreted and prove useful later when considering the final result at TDL. Our benchmark not only highlights the challenges in performing periodic quantum chemistry -- reflecting the current immaturity of the field -- but also confirms that our chosen basis sets are adequate. In the main text, we discuss in detail how these errors can be accounted for in our final many-body predictions.
\clearpage

\subsection{RPA adsorption energy for the CO@Cu structure at thermodynamic limit}

\begin{figure*}[ht]
    \centering
    \includegraphics[width=0.85\linewidth]{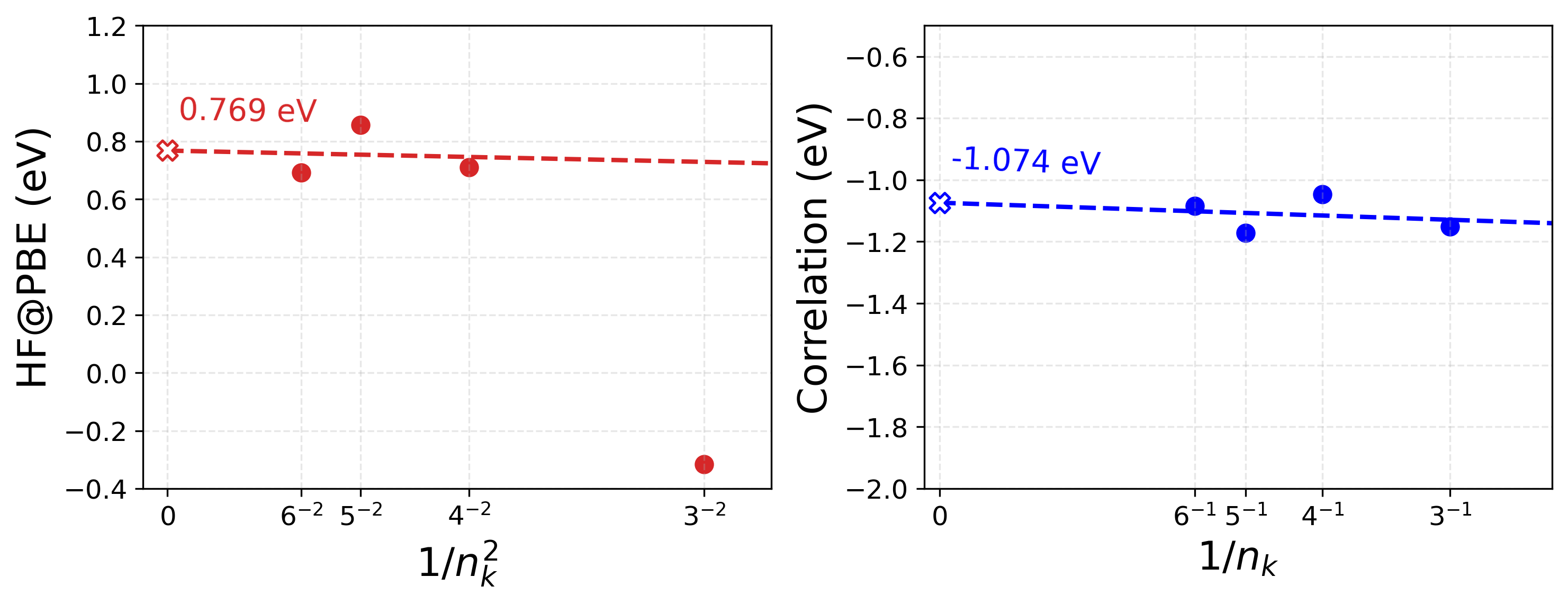}
    \caption{
    Convergence of the adsorption energy contributions with respect to $k$-point sampling for the atop structure, using RPA. The left panel shows the Hartree--Fock (HF@PBE) contribution to the adsorption energy, obtained by evaluating the HF energy on DFT orbitals. The right panel shows the corresponding RPA correlation contribution. Each panel plots energy as a function of inverse squared or inverse $k$-point density, respectively, with dashed lines representing linear extrapolations to the thermodynamic limit ($1/n_k^2 \to 0$ or $1/n_k \to 0$). The extrapolated values at $1/n_k^2 = 0$ or $1/n_k = 0$ are highlighted and annotated. The HF term converges more slowly and exhibits greater variation, while the correlation contribution displays smoother and more stable convergence behavior.
    }
  \label{fig:rpa_top}
\end{figure*}

\begin{figure*}[ht]
      \centering
      \includegraphics[width=0.85\linewidth]{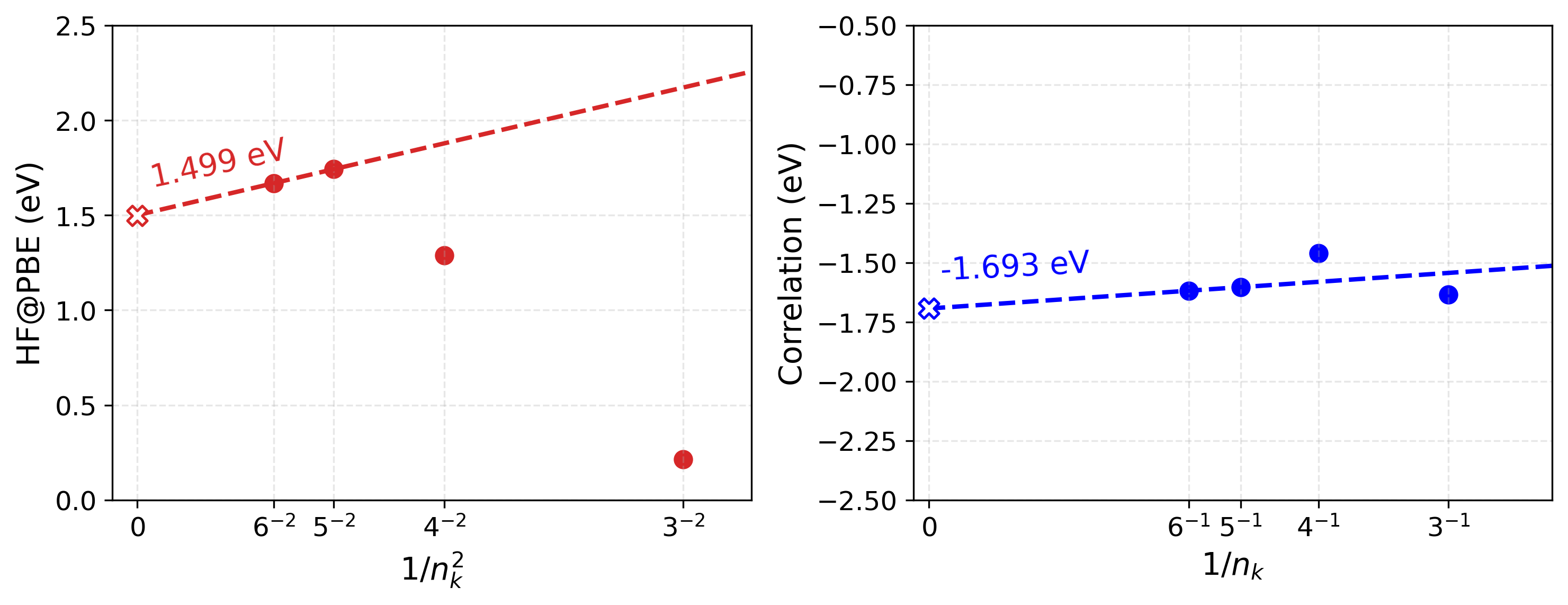}
      \caption{Same as Fig.~\ref{fig:rpa_top} but for the fcc structure}
      \label{fig:fermion}
\end{figure*}
\clearpage

\subsection{CO@Cu results from literature and current work}

\begin{table}[h!t]
    \centering
    \begin{tabular}{cccccc}
    \hline
         &  &  Atop&  Fcc&  Gap& Ref. \\
    \hline
    \multirow{3}{*}{LDA} 
         &  LDA&  -1.29 &  -1.66 &  -0.37 
& \citenum{stroppaCOAdsorption2007}\\
         &  LDA&  -1.58 &  -2.01 &  -0.43 & \citenum{NEEF20061085} \\
         &  LDA&  -1.16&  -1.47&  -0.31& \citenum{hsingQuantumMonte2019}\\
    \hline
    \multirow{8}{*}{GGA} 
         &  PBE&  -0.71 &  -0.87 &  -0.17 & \citenum{stroppaCOAdsorption2007} \\
         &  PBE&  -0.61&  -0.71&  -0.10& \citenum{hsingQuantumMonte2019}\\
         &  PBE&  -0.75&  -0.85&  -0.10& \citenum{huExactTreatment2007} \\
         &  PBE+D3BJ&  -0.99&  -1.09&  -0.10& \citenum{huExactTreatment2007}\\
         &  RPBE&  -0.28 &  -0.39 &  -0.11 & \citenum{gameelUnveilingCO2018}\\
         &  M06-L&  -0.71&  -0.75&  -0.04& \citenum{huExactTreatment2007}\\
         & PW91& -1.06 & -1.31 & -0.25 &\citenum{NEEF20061085}\\
         & PBE& -0.79& -0.92& -0.13&This work\\
    \hline
    \multirow{2}{*}{DFT + XC corr} 
         & GGA& & & 0.01&\citenum{huExactTreatment2007}\\
         & B3LYP& & & 0.21&\citenum{huExactTreatment2007}\\
    \hline
    \multirow{6}{*}{Hybrid} 
 & PBE0& -0.61& -0.87& -0.27&\citenum{stroppaCOAdsorption2007}\\
 & HSE03& -0.56& -0.56& 0.01&\citenum{stroppaCOAdsorption2007}\\
 & B3LYP& -0.57& -0.44& 0.13&\citenum{NEEF20061085} \\
 & B3LYP& -0.27& 0.16& 0.43&\citenum{huExactTreatment2007}\\
 & B3LYP+D3BJ& -0.69& -0.26& 0.43&\citenum{huExactTreatment2007}\\
 & PBE0& -0.62& -0.53& 0.09&This work\\
    \hline
    \multirow{6}{*}{RPA} 
 & RPA@LDA& -0.34& -0.18& 0.16&\citenum{renExploringRandom2009} \\
 & RPA@PBE& -0.35& -0.17& 0.18&\citenum{renExploringRandom2009} \\
 & RPA@PBE0& -0.37& -0.15& 0.22&\citenum{renExploringRandom2009} \\
 & RPA@PBE& -0.42& -0.31& 0.11&\citenum{Schimka2010_mbpt_surface_adsorption}\\
 & RPA@PBE (4x4x1)& -0.34& -0.17& 0.17&This work\\
 & RPA@PBE (TDL)& -0.31& -0.19& 0.11&This work\\
    \hline
 QMC& DMC& -0.65& -0.25& 0.40& \citenum{hsingQuantumMonte2019}\\
    \hline
    \multirow{3}{*}{Embed}
& XYG3& -0.63& -0.51& 0.12&\citenum{chenAccurateDescriptions2023}\\
 & MRSDCI& -0.49& & &\citenum{sharifzadehEmbeddedConfiguration2008}\\
 & AFQMC@RPA& -0.52& -0.25& 0.28&This work\\
    \hline
    \multirow{2}{*}{Experiment} 
 & &-0.52 $\sim$ -0.45 & & &\citenum{vollmerDeterminationSite2001,hollinsInteractionsCO1979,kesslerChemisorptionCO1977,pritchardStructureCO1979,bartelsEvolutionCO1999,kirsteinCOAdsorption1986}\\
 && -0.59 & & &\citenum{wellendorffBenchmarkDatabase2015}\\
    \hline
    \end{tabular}
    \caption{Data from literature and current work for CO adsorption on Cu(111).
Comparison of adsorption energies (in eV) for the \textit{atop} and \textit{fcc} sites, and the energy gap between them (Gap = $E_{fcc}-E_{atop}$). A positive value means adsorption on \textit{atop} site favored than that on \textit{fcc} site).}
    \label{tab:COCu_data}
\end{table}
\clearpage

\subsection{DFT cross-validation of GTO and planewave methods for \ce{H2}@Cu(111)}

\begin{table}
    \centering
    \begin{tabular}{ccccc}
    \hline
& \multicolumn{2}{c}{PW} & \multicolumn{2}{c}{GTO}\\
\cline{2-5}
 kmesh& 3L& 5L& 3L& 5L\\
  \cline{2-5}
       2$\times$2$\times$1  &  0.631 &  0.560&  0.559&  0.704\\
       3$\times$3$\times$1  &  0.694 &  0.669&  0.667&  0.658\\
       4$\times$4$\times$1  &  0.570 &  0.575&  0.581&  0.560\\
       5$\times$5$\times$1  &  0.521 &  0.541&  0.492&  0.522\\
       6$\times$6$\times$1  &  0.557 &  0.578&  0.584&  0.604\\
       \hline
    \end{tabular}
    \caption{Comparison between computed energy barrier (in eV) obtained using different methods. Results from PySCF with a Gaussian-type orbital (GTO) basis are compared with those from VASP using a plane-wave basis set at various k-point grids.}
    \label{tab:ene_pyscf_vasp_h2}
\end{table}
To validate our computational model and parameters, we benchmarked the \ce{H2} desorption energy barrier calculated with the PBE functional against reference plane-wave calculations from VASP. As shown in Table~\ref{tab:ene_pyscf_vasp_h2}, our results are in good agreement with the VASP reference, with the deviation in the desorption barrier falling within chemical accuracy (1 kcal/mol = 0.043 eV).

This agreement confirms that our 3-layer slab model provides results comparable to a more extensive 5-layer model. Furthermore, it validates the use of a 2$\times$2$\times$1 k-mesh for this system, as this setup successfully reproduces the values obtained with denser k-point sampling in the plane-wave code. Therefore, the chosen parameters are considered appropriate for studying this system.
\clearpage

\subsection{\ce{H2}@Cu results with extended fragment size}

\begin{table}
    \centering
    \begin{tabular}{ccccl}\hline
         Fragment&  $\Delta E_{corr}$/eV&  $\sigma$/eV& $E_{{barrier}}$/eV \\\hline
         $\ce{H2}$&  -0.171&  0.005& 0.669   \\
         $\ce{H2Cu2}$&  -0.081&  0.046& 0.759\\
         $\ce{H2Cu4}$&  0.029&  0.067& 0.869  \\ 
         $\ce{H2Cu6}$&  0.017&  0.094& 0.857 \\ 
         \hline
    \end{tabular}
    \caption{The calculated \ce{H2} desorption barrier with respect to the fragment size in the FEMION embedding framework. The correlation energy correction, $\Delta E_{\text{corr}}$, is defined as the difference between the fragment's AFQMC and RPA energies ($E_{\text{f}}^{\text{AFQMC}} - E_{\text{f}}^{\text{RPA}}$), with $\sigma$ representing the associated statistical error of the stochastic AFQMC calculation. $E_{barrier}(ewald)$ and $E_{barrier}$ denote the reaction barrier after correction by FEMION with and without the Ewald correction for exchange-divergence, respectively. We systematically enlarge the fragment by including Cu atoms from the first surface layer. The barrier energy stabilizes at the $\ce{H2Cu4}$ level; because further increasing the fragment size to $\ce{H2Cu6}$ increases the statistical error ($\sigma$) without significantly changing the energy, we adopt $\ce{H2Cu4}$ as the final result reported in the main text. For metallic systems, it has been suggested that the finite-size errors arising from the exact exchange and the post-meanfield correlation energy tend to cancel each other out~\cite{harlAssessingQualityRandom2010}. Following this reasoning, we adopt the uncorrected barrier, $E_{{barrier}}$, as our final and most physically relevant result for this system.} 
    \label{tab:emb_h2}
\end{table}
\clearpage
\subsection{\ce{H2}@Cu results from literature and current work}
\begin{table}[h]
    \centering
    \begin{tabular}{cccc}
        \hline
        & Method & $E_{barrier}/\rm{eV}$ & Ref. \\
        \hline
        \multirow{3}{*}{DFT} 
        &PBE+D3BJ & 0.58 & \citenum{zhaoBenchmarkingEmbedded2020} \\
        &PBE+D3BJ & 0.55 & \citenum{weiIntroducingEmbedded2023} \\
        &PBE & 0.55 & This work \\
        \hline
        \multirow{5}{*}{RPA}
        &RPA@PBE & 0.59 & 
        \citenum{weiIntroducingEmbedded2023} \\
        &RPA@PBE (1-layer) & 0.72 & 
        \citenum{weiIntroducingEmbedded2023} \\
        &RPA@PBE (2-layer) & 0.66 & \citenum{weiIntroducingEmbedded2023} \\
        &RPA@PBE (3-layer) & 0.60 & \citenum{weiIntroducingEmbedded2023} \\
        &RPA@PBE & 0.83 & This work \\
        \hline
        \multirow{2}{*}{Cluster-emb} 
        &B3LYP@PBE+D3BJ & 0.94 & \citenum{chenAccurateDescriptions2023} \\
        &XYG3@PBE & 0.88 & \citenum{chenAccurateDescriptions2023} \\
        \hline
        \multirow{6}{*}{PBC-emb} 
        &CASPT2 & 1.00 & \citenum{zhaoBenchmarkingEmbedded2020} \\
        &CASPT2 no 3s3p & 0.75 & \citenum{zhaoBenchmarkingEmbedded2020} \\
        &NEVPT2 & 0.74 & \citenum{zhaoBenchmarkingEmbedded2020} \\
        &NEVPT2 no 3s3p & 0.69 & \citenum{zhaoBenchmarkingEmbedded2020} \\
        &ASCI-SCF-PT2 & 1.04 & \citenum{zhaoBenchmarkingEmbedded2020} \\
        &ASCI-SCF-PT2+MP2 & 1.32 & \citenum{zhaoBenchmarkingEmbedded2020} \\
        &AFQMC@RPA & 0.82 & This work \\
        \hline
        \multirow{1}{*}{Experiment} 
        & & 0.84 & \citenum{chenAccurateDescriptions2023,caoHydrogenAdsorption2018} \\
        \hline
    \end{tabular}
    \caption{Data from literature and current work for \ce{H2} desorption from Cu(111). The energy barrier for the desorption is list in table in unit of eV. The experimental value is corrected for ZPM using the correction obtained from Ref.~\citenum{chenAccurateDescriptions2023} via the PBE functional~\cite{PBE}. }
    \label{tab:H2Cu_data}
\end{table}
\clearpage

\subsection{Impact of Thermal Smearing on Embedding Energies}
\begin{figure*}[ht]
    \centering
    \includegraphics[width=0.85\linewidth]{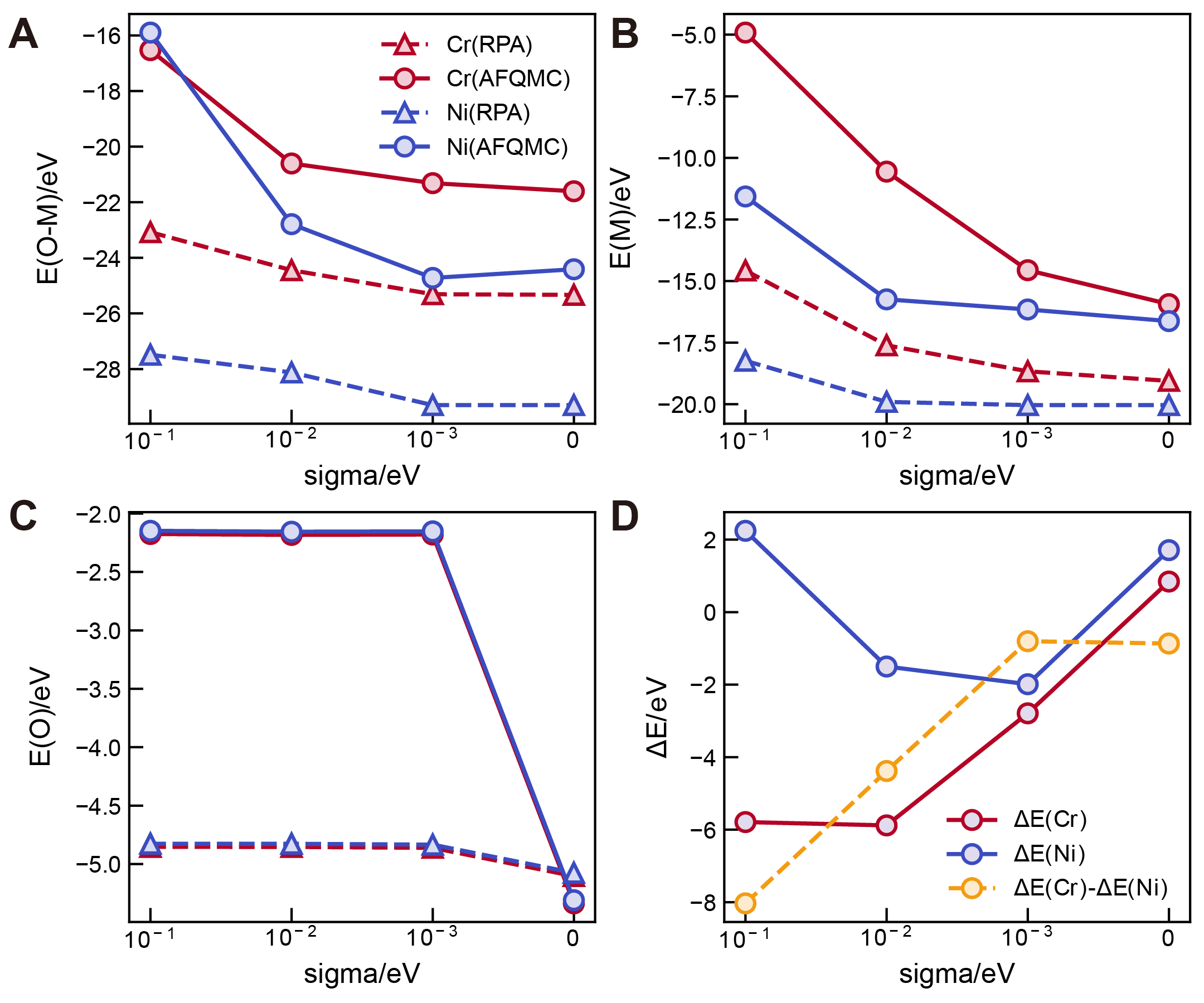}
    \caption{Thermal smearing parameter test on O@Cr-Cu(111) and O@Ni-Cu(111). (A-C) Total energy of the fragment in the adsorbed structure, slab structure, and adsorbate structure, respectively, as a function of the smearing parameter~$\sigma$. (D) The embedding energy correction, where $\rm{\Delta E=E(O-M)-E(M)-E(O)}$; highlights the difference between the corrections for Cr and Ni. Note that the large unphysical difference vanishes as $\sigma$ to 0.}
  \label{fig:smear_test}
\end{figure*}
However, while thermal smearing with fractional occupations is commonly employed to improve numerical stability in mean-field calculations of metallic systems, care must be taken to ensure that the resulting energies are properly extrapolated to the zero-temperature (ground-state) limit. This requirement becomes particularly critical in high-precision post-mean-field and embedding theories, where residual finite-temperature contributions can lead to spurious energy differences.

To validate the robustness of our predictions with respect to the smearing parameter, we performed systematic convergence tests on the smearing width ($\sigma$) across a range of adsorption species. Figure~\ref{fig:smear_test} presents the results for O@Cr-Cu(111) and O@Ni-Cu(111) as examples to illustrate the sensitivity of the energy components. We observed that this sensitivity varies significantly across the transition metal series, with Cr-based systems showing a stronger dependence on the smearing width than Ni-based systems. Consequently, at $\sigma = 0.1$ eV, this differential sensitivity introduces a large, artificial energy discrepancy between the two metals (Figure~\ref{fig:smear_test}D).

Crucially, as $\sigma \rightarrow 0$, this unphysical fluctuation vanishes, and the embedding energy corrections for both metals converge to consistent, small values. This confirms that the discrepancy observed at finite smearing originates from a finite-temperatures artifact, rather than a physical property.

To eliminate these artifacts and ensure robustness of the predicted ground-state energies with respect to the smearing parameter, we implemented a zero-temperature occupation scheme in the FEMION calculations. Since strictly setting $\sigma=0$ can lead to convergence issues due to high orbital degeneracy near the Fermi level, our refined scheme treats orbitals with energy differences below a threshold of $10^{-6}$ Hartree ($\sim 2.7 \times 10^{-5}$ eV) as degenerate. This allows for fractional occupation only within a controlled near-degenerate subspace, ensuring robust convergence without introducing broad thermal smearing errors.
\clearpage

\subsection{Validation against DFT Hierarchy}
\begin{figure*}[ht]
    \centering
    \includegraphics[width=0.65\linewidth]{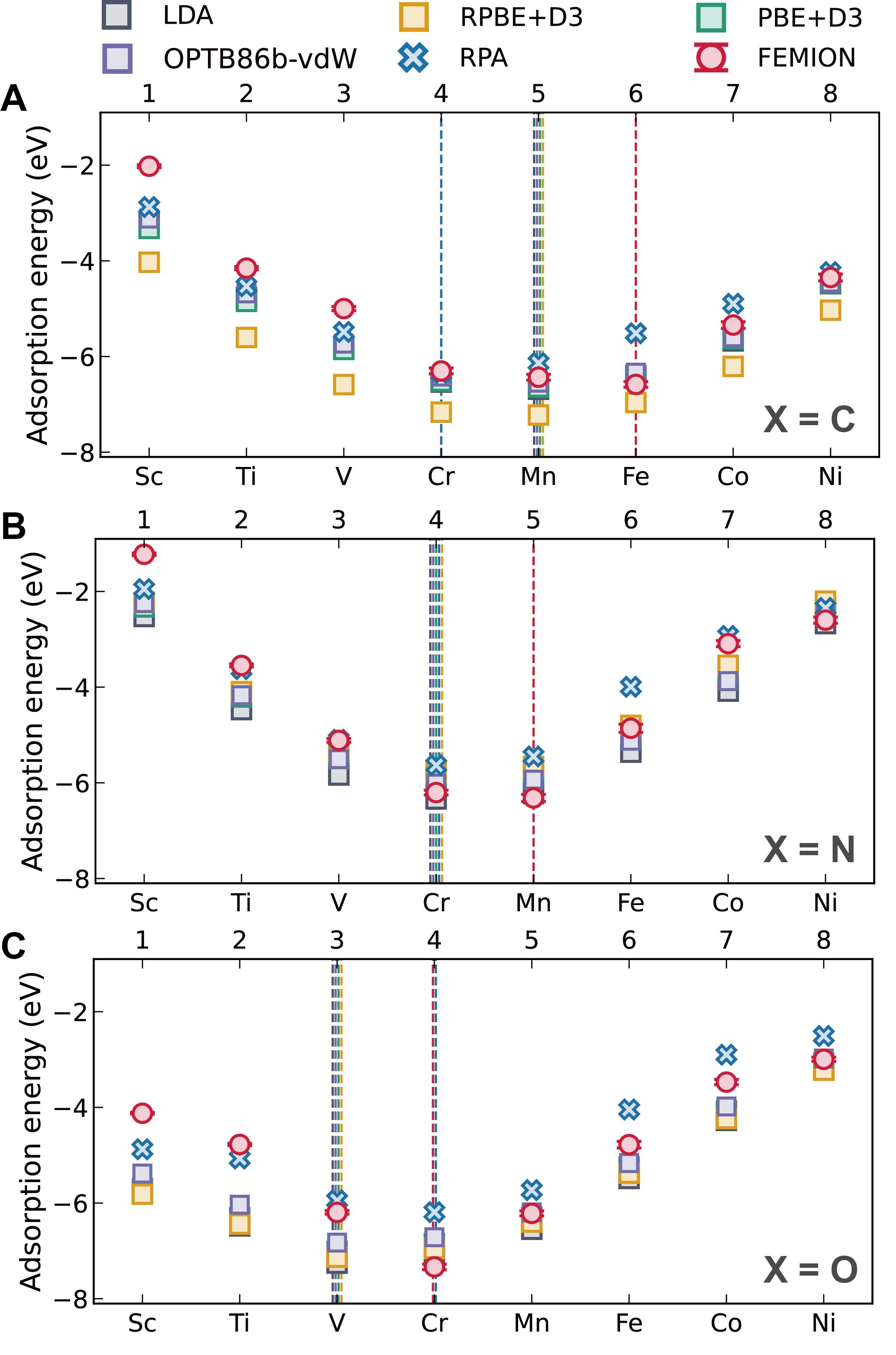}
    \caption{Benchmark of adsorption energies for (A) C, (B) N, and (C) O across the 3d transition metal series (Sc–Ni). The plots compare the results obtained from the AFQMC solver (FEMION, red circles) and RPA (blue crosses) against a hierarchy of DFT functionals (LDA, PBE+D3, RPBE+D3, and OPTB86b-vdW). Computational settings are detailed in~\ref{subsec:SAA}. }
  \label{fig:DFT_range}
\end{figure*}

While the direct experimental benchmarks for these specific transition-metal doped single atom alloys are absent, we compared the FEMION (RPA+AFQMC) results against a hierarchy of DFT functionals (LDA~\cite{LDA}, PBE+D3~\cite{PBE}, RPBE+D3~\cite{RPBE}, optB86b-vdW~\cite{klimesChemicalAccuracyVan2009,klimesVanWaalsDensity2011}) and RPA to ensure the reliability of our predictions. To account for long-range van der Waals interactions, the DFT-D3~\cite{DFTD3} method with Becke–Johnson (BJ) damping~\cite{D3BJ} was applied for PBE and RPBE.
As shown in Figure~\ref{fig:DFT_range}, the FEMION adsorption energies consistently fall within the reference ranges established by these functionals, confirming that our method yields physically robust energy magnitudes. 
Notably, all DFT functionals predicts a biased preference for the most stable adsorption element, in disagreement with the 10-electron count rule. RPA agrees on the O adsorption, and FEMION agrees on all C, N and O adsorptions. 
However, a critical divergence arises regarding the qualitative trend. All tested DFT functionals consistently predict a biased preference for the most stable adsorption element, contradicting the 10-electron count rule. While RPA partially corrects this deviation by recovering the expected behavior for O, only FEMION achieves consistent agreement with the 10-electron count rule across all adsorbates (C, N, and O). This highlights the importance of including more advanced electron correlation treatment beyond GGA-DFT. 

\clearpage

\bibliography{refs}